\documentclass[12pt, onecolumn, draftclsnofoot, a4paper]{article}
\usepackage{graphicx}
\usepackage[caption=false,font=footnotesize]{subfig}
\usepackage{multirow}
\usepackage{array}
\usepackage[noadjust]{cite}
\usepackage{url}
\usepackage{diagbox}
\usepackage{mathtools}
\usepackage{amsmath}
\usepackage{amssymb}
\usepackage{empheq}
\usepackage{bm}
\usepackage{color}
\usepackage{gensymb}
\usepackage{algorithm}
\usepackage{algpseudocode}
\usepackage{amsthm}
\usepackage[many]{tcolorbox}
\usepackage[shortlabels]{enumitem}
\usepackage{booktabs}
\usepackage[margin=0.59in]{geometry}

\theoremstyle{definition}

\newtcolorbox{framefloat}[1][]{fonttitle=\bfseries, titlerule=0pt, boxrule=0.5pt,
  colframe=black,colback=white,float=!t,#1}

\newcommand\blfootnote[1]{%
  \begingroup
  \renewcommand\thefootnote{}\footnote{#1}%
  \addtocounter{footnote}{-1}%
  \endgroup
}

\definecolor{MyRed}{RGB}{134, 8, 8}

\allowdisplaybreaks

\usepackage[makeroom]{cancel}
\makeatletter
\def\cantox@vector#1#2#3#4#5#6#7#8{%
  \dimen@.5\p@
  \setbox\z@\vbox{\boxmaxdepth.5\p@
   \hbox{\kern-1.2\p@\kern#1\dimen@$#7{#8}\m@th$}}%
  \ifx\canto@fil\hidewidth  \wd\z@\z@ \else \kern-#6\unitlength \fi
  \ooalign{%
    \canto@fil$\m@th \CancelColor
    \vcenter{\hbox{\dimen@#6\unitlength \kern\dimen@
      \multiply\dimen@#4\divide\dimen@#3 \vrule\@depth\dimen@\@width\z@
      \vector(#3,-#4){#5}%
    }}_{\raise-#2\dimen@\copy\z@\kern-\scriptspace}$%
    \canto@fil \cr
    \hfil \box\@tempboxa \kern\wd\z@ \hfil \cr}}
\def\bcancelto#1#2{\let\canto@vector\cantox@vector\cancelto{#1}{#2}}
\makeatother

\makeatletter
\newcommand*\rel@kern[1]{\kern#1\dimexpr\macc@kerna}
\newcommand*\widebar[1]{%
  \begingroup
  \def\mathaccent##1##2{%
    \rel@kern{0.8}%
    \overline{\rel@kern{-0.8}\macc@nucleus\rel@kern{0.2}}%
    \rel@kern{-0.2}%
  }%
  \macc@depth\@ne
  \let\math@bgroup\@empty \let\math@egroup\macc@set@skewchar
  \mathsurround\z@ \frozen@everymath{\mathgroup\macc@group\relax}%
  \macc@set@skewchar\relax
  \let\mathaccentV\macc@nested@a
  \macc@nested@a\relax111{#1}%
  \endgroup
}
\makeatother

\providecommand{\keywords}[1]
{
  \small	
  \textbf{\textit{Keywords---}} #1
}

\begin{document}

\title{Probabilistically Robust Optimization of IRS-aided SWIPT Under Coordinated Spectrum Underlay}
\author{Konstantinos Ntougias and Ioannis Krikidis}
\date{}

\maketitle

\begin{abstract}
This study considers the Joint Transmit/Reflect Beamforming and Power Splitting (JTRBPS) optimization problem in a spectrum underlay setting, such that the transmit sum-energy of the intelligent reflecting surface (IRS)-aided secondary transmitter (ST) is minimized subject to the quality-of-service requirements of the PS-simultaneous wireless information and power transfer (SWIPT) secondary receivers and the interference constraints of the primary receivers (PR). The interference at the PRs caused by the reception of IRS-reflected signals sent by the primary transmitter is taken into account. A coordinated channel state information (CSI) acquisition protocol is proposed. Next, assuming availability at the ST of perfect CSI for all direct and IRS-cascaded transmitter--receiver channels, two penalty-based iterative algorithms are developed: an alternating minimization algorithm that involves semi-definite relaxation in JTBPS design and successive convex approximation in RB optimization, and a block coordinate descent algorithm that employs the Riemannian conjugate gradient algorithm in RB updates. Finally, an outage-constrained robust design under imperfect CSI is devised. Numerical simulations highlight the performance gains of the proposed strategies over benchmarks, corroborate the benefits of using an IRS, and provide valuable insights.
\end{abstract}


\section{Introduction}\label{sec:1}
\blfootnote{This work has received funding from the European Research Council (ERC) under the European Union's Horizon 2020 research and innovation program (Grant agreement No. 819819). 

The authors are with the Department of Electrical and Computer Engineering, University of Cyprus, 1678 Nicosia, Cyprus. (E-mail: \{ntougias.konstantinos, krikidis\}@ucy.ac.cy).}
The commercialization of 5th Generation (5G) networks has been followed by an ongoing proliferation of terminals, such as smartphones and Internet-of-Things (IoT) devices, as well as by a rapid growth of the mobile data traffic, which is mainly driven by video delivery applications~\cite{EMR2021}. Furthermore, emerging use cases such as Industrial IoT (IIoT) ``push'' the execution of machine learning algorithms to IoT nodes. 

Energy-demanding tasks, such as video streaming and learning, can quickly drain the battery of the user terminals. Simultaneous wireless information and power transfer (SWIPT) refers to a recently proposed paradigm that integrates wireless power supply with downlink (DL) communication to prolong the limited lifetime of such energy-constrained devices~\cite{ClerckxxSWIPTsurv2019}. Massive multiple-input multiple-output (MIMO) technology, in turn, is widely applied as a means to enhance the spectral efficiency (SE) through aggressive spatial multiplexing~\cite{mMIMOBook}. 

The virtual extension of the usable bandwidth through spectrum sharing constitutes yet another approach for coping with the traffic growth under the evident spectral scarcity~\cite{SSVerHor}. This spectrum usage model is also promoted as a strategy for fast and low-cost deployment of private IIoT networks, which have risen in response to the growing demands of verticals for increased network control~\cite{IIoT}. Concurrent access to the shared spectrum, also known as spectrum underlay, represents the most efficient variant, in terms of resources utilization. Regardless of its ``flavor'', the adopted spectrum sharing scheme requires an interference management mechanism to ensure the provision of quality-of-service (QoS) guarantees to the involved parties. This is typically accomplished nowadays through some inter-system coordination scheme~\cite{IIoT,CBRS,LSA,Boccardi}.  

It becomes apparent that recent trends urge us to combine the SWIPT and coordinated spectrum underlay paradigms, in order to achieve energy-autonomous and spectrally-efficient network operation. Massive MIMO can realize this vision by effectively suppressing the co-channel interference (CCI) and compensating for the propagation loss that deteriorates the receive signal-to-noise-ratio (SNR) and hampers the efficiency of radio frequency (RF) energy harvesting (EH) via transmit beamforming (TB)~\cite{mMIMOSS,mMIMOSWIPT3}. Nonetheless, the installation of an excessive number of energy-hungry and expensive RF units at the access points (AP) is particularly problematic in dense network setups as well as in private IIoT networks, which are commonly subject to (s.t.) strict operational constraints.  

The disruptive intelligent reflecting surface (IRS) technology utilizes passive reflect beamforming (RB) instead of active TB to address these issues~\cite{IRSSurvey}. Specifically, this intermediary is equipped with a large number of passive reflecting elements that independently scatter each any incident radio signal with a controllable amplitude attenuation or/and phase shift. By jointly adjusting all reflection coefficients\footnote{The joint optimization of both the reflection amplitudes and phase shifts of all IRS elements results in high implementation cost, especially for certain IRS hardware realizations, and algorithmic complexity~\cite{IRSSurvey}. Therefore, the vast majority of works on joint TB/RB optimization in the literature assumes that the reflection amplitudes are set to unity, ignoring both the reflection loss which is small and their dependency on the phase shifts, such that the strength of the received signals at the indented users is maximized and solely focuses on the design of the IRS phase shifts matrix to achieve a desirable performance--complexity trade-off. Moreover, a substantial portion of these works considers continuous phase shifts, although hardware limitations and cost constraints lead to discrete phase shift implementations~\cite{IRSSurvey}, to further reduce the algorithmic complexity and encourage practical application. These low-complexity algorithms provide performance benchmarks and facilitate the design of variants that achieve a significant fraction of these performance limits in practice. Consequently, we follow these approaches as well.}, we can boost the receive SNR at target users or/and mitigate the CCI at other directions in a cost-effective and energy-efficient manner. 

\subsection{Related Work}\label{subsec:1.1}
Joint TB/RB optimization has been studied in the literature under a variety of IRS-aided wireless communication settings and scenarios. The work~\cite{IRSRayleigh} investigates SE maximization for an IRS-assisted point-to-point (PTP) multiple-input single-output (MISO) link, while~\cite{IRSSec} assumes coexistence with a group of single-antenna eavesdroppers and develops a robust algorithm that minimizes the transmit sum-power under imperfect channel state information (CSI) for the IRS-cascaded link with the legitimate user. On the other hand,~\cite{IRSPhySec1,IRSPhySec3} tackle the maximization of the secrecy rate in a PTP MISO or MIMO link that is collocated with a single- or multi-antenna eavesdropper, respectively. The study~\cite{IRSMulticast1}, in turn, explores sum-rate maximization in a multigroup multicast setup.  

Other works consider an IRS-enhanced MISO broadcast setup. Sum-rate and energy efficiency (EE) maximization is investigated in~\cite{IRSPwr,IRSEE}, respectively, under the application of zero-forcing (ZF) precoding, whereas~\cite{IRS1,IRS3,Provable} are devoted to transmit sum-power minimization assuming continuous or discrete IRS phase shifts or a system with multiple IRSs, respectively. The work~\cite{IRSWSRmax} studies weighted sum-rate (WSR) maximization in setups with discrete IRS phase shifts, while~\cite{IRSAmplitude} explores robust WSR maximization assuming imperfect CSI and control over the reflection amplitudes. The study~\cite{IRSImpCSI1} proposes a worst-case design for transmit sum-power minimization under a scenario where only the IRS--user reflection channels are imperfectly known, whereas~\cite{IRSCascCSI1} focuses on worst-case and outage-constrained robust designs under imperfect CSI of either the IRS-cascaded or the effective AP--user links. In~\cite{IRSImpCSI2}, the authors develop a robust algorithm that maximizes the sum-rate under imperfect CSI, assuming a setup with multiple multi-antenna eavesdroppers.

\subsection{Motivation and Objectives}\label{subsec:1.2} 
The work~\cite{WCL2} deals with the multi-objective optimization problem of maximizing the sum-rate and the total harvested energy in an IRS-aided multi-user MISO SWIPT setup with separate information decoding (ID) and EH receivers, whereas in~\cite{IRSSWIPT0,IRSSWIPT1,IRSSWIPT2} the authors develop algorithms for maximizing the minimum power or the weighted sum-power at the EH receivers or for minimizing the transmit sum-power, respectively. The latter problem is revisited in~\cite{IRSSWIPTImpCSI} assuming power-splitting (PS)-SWIPT receivers and imperfect CSI, while~\cite{Proposed1} studies the WSR maximization problem assuming a MIMO broadcasting setup for PS-SWIPT. Finally,~\cite{WCL1} proposes a scheme that maximizes a utility function which balances the EE of information and energy transmission and~\cite{Provable2} tackles the max-min EE fairness problem. 

Rate maximization for a secondary IRS-assisted PTP single-input single-output (SISO) or MISO link that is collocated with one or multiple single-antenna primary receivers (PR), respectively, is studied in~\cite{IRSSS,IRSCR}. The authors in~\cite{IRSCRRobust} consider a secondary IRS-enhanced multi-user MISO DL system that is collocated with a single-antenna PR and derive a robust design assuming imperfect CSI regarding the effective secondary transmitter (ST)--PR link to minimize the transmit sum-power of the ST. Sum-rate maximization of an IRS-enabled MISO broadcast system that coexists with a primary multi-user MISO DL setup is investigated in~\cite{SchoberCR}.

Surprisingly enough, the design of secondary IRS-aided multi-user SWIPT systems that operate in a spectrum underlay regime, in accordance with the earlier discussion, has not been studied yet, to the best of our knowledge. The main objective of this work is to fill in this gap in the literature and highlight the interplay between the QoS constraints of the secondary receivers (SR) and the interference energy constraints of the PRs as well as the dual role of the IRS as a means to meet both types of requirements. 

\subsection{Contributions}\label{subsec:1.3}
\begin{figure}[!t]
\centering
\captionsetup{justification=centering}
\includegraphics[scale = 0.7]{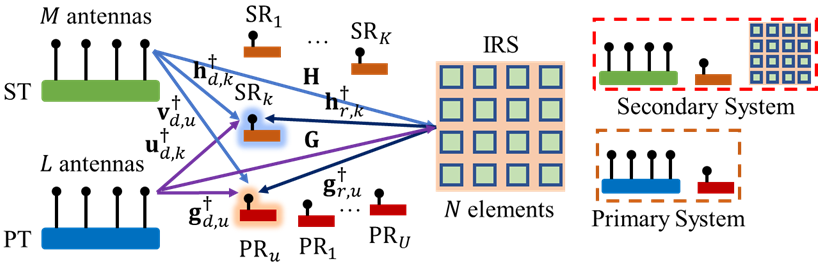}
\caption{System setup and channels.}
\label{fig:1}
\end{figure}

We consider a spectrum underlay setup comprised by a primary and a secondary MISO broadcast system, wherein an IRS assists the transmissions of the ST and each SR adopts the PS-SWIPT architecture. Each transmitter is linked with each receiver through both a direct and an IRS-cascaded, intra- or inter-system channel, as shown in Fig.~\ref{fig:1}. We study the joint TB/RB and receive PS (JTRBPS) optimization problem, such that the transmit sum-energy of the ST is minimized s.t. the QoS constraints of the SRs and the interference energy constraints of the PRs. Our goal is to derive \textit{low-complexity, near-optimal designs that can be applied in practice}.

Besides the differences of this paper with relevant studies in terms of the considered setup and problem, \textit{this is the first work that investigates the combination of IRS-aided SWIPT with IRS-assisted spectrum underlay}, to the best of the authors' knowledge. Let us present some notable characteristics of this study:
\begin{itemize}
\item The considered setup is generic and encompasses systems with separated ID and EH receivers~\cite{WCL2,IRSSWIPT0,IRSSWIPT1,IRSSWIPT2} or with a single SR~\cite{IRSCR} or PR~\cite{IRSCRRobust} as special cases.
\item A non-linear EH model is utilized, as opposed to the innacurate linear EH model~\cite{WCL2,IRSSWIPT0,IRSSWIPT1,IRSSWIPT2}. 
\item The reverse inter-system interference (RISI) at the SRs, which is attributed to the transmissions of the PT and is often neglected in relevant studies~\cite{IRSSS,IRSCR,IRSCRRobust,SchoberCR}, is taken into account in system design and performance evaluation. The RISI degrades the signal-to-interference-plus-noise-ratio (SINR) and enhances the harvested direct-current (DC) energy of the SRs.
\item The PT serves its users selfishly, i.e., it designs its TB vectors based on its direct channels with the PRs as if the primary system were isolated, as it is typically the case in spectrum underlay setups. Therefore, in contrast to prior works~\cite{IRSSS,IRSCR,IRSCRRobust,SchoberCR}, our designs consider, on top of the forward ISI (FISI) at the PRs incurred by the ST, the intra-primary-system interference caused by the reception at the PRs of interfering and useful data signals sent by the PT through the respective IRS-cascaded channels.
\item As in many previous works, we consider unit reflection amplitudes and continuous IRS phase shifts to reduce the algorithmic complexity. Nevertheless, we map the optimal phase shifts to the closest possible discrete values utilized in practice to quantify the performance degradation imposed by this design approach on actual implementations. This is often ignored in the literature.
\end{itemize}

The individual contributions of this work are listed next: 
\begin{itemize}
\item An inter-system coordination protocol that enables side information acquisition at the ST to facilitate interference management via TB or/and RB is specified, as opposed to relevant works~\cite{IRSSS,IRSCR,IRSCRRobust,SchoberCR}. We also consider the performance impact of the coordination overhead.  
\item Assuming initially availability of perfect CSI regarding all effective transmitter-receiver links at the ST, we develop two penalty-based iterative algorithms to tackle the challenging non-convex optimization problem of interest. The first one is an alternating minimization (AM) algorithm that involves the semi-definite relaxation (SDR) method to solve the JTBPS problem and the successive convex approximation (SCA) technique to enforce a feasible rank-one RB solution and guarantee convergence~\cite{Provable,Provable2}, in contrast to the standard SDR/Gaussian randomization (GR) approach.
\item The second algorithm is based on the decoupling of the optimization variables via problem reformulation and makes use of the block coordinate descent (BCD) framework. In this case, the non-convex unit modulus constraints of the RB matrix are handled via the Riemannian conjugate gradient (RCG) algorithm~\cite{IRSRayleigh}.
\item CSI imperfections are inevitable in practice, thus rendering robust designs imperative. This is more emphatic in IRS-aided setups, where the channel estimation errors of the direct channels contaminate the cascaded CSI~\cite{Proposed2}. However, prior studies on IRS-aided SWIPT or IRS-assisted spectrum underlay either assume that only the reflected IRS--user channels~\cite{IRSSWIPTImpCSI} or the IRS-cascaded ST-PR links~\cite{IRSCRRobust} are error-prone or focus on worst-case robust designs~\cite{IRSCR,IRSSWIPTImpCSI} which are conservative, since they consider extreme yet rare channel conditions. On the contrary, we develop an outage-constrained robust design under the realistic assumption of imperfect CSI for both the direct and IRS-cascaded links of each transmitter with each receiver. To this end, we extend the AM-based scheme by adopting a statistical CSI error model to capture the channel uncertainty and Bernstein type inequalities (BTI) to obtain convex approximations of the constraints~\cite{RobustOpt}. Also, the penalty convex-concave procedure (CCP) method is employed to handle the non-convex unit modulus constraints of the RB matrix.
\item We study the convergence and complexity of the proposed algorithms.
\end{itemize}

Extensive numerical simulations highlight the performance gains of the proposed schemes over benchmarks, corroborate the beneficial role of the IRS, and provide valuable insights.
\begin{table}[!t]
\centering 
\caption{Abbreviations}
\label{tab:tab1}
\resizebox{\textwidth}{!}{%
\begin{tabular}{@{}llll@{}}
\toprule
\textbf{Abbreviation} & \textbf{Definition} & \textbf{Abbreviation} & \textbf{Definition}                           \\ 
\midrule
AM 	   & Alternating Minimization 	& MISO		& Multiple-Input Single-Output \\
BCD    & Block Coordinate Descent	& PR/PT		& Primary Receiver/Transmitter \\
BTI    & Bernstein-Type Inequality	& PSD    	& Positive Semi-Definite \\
CCP    & Convex-Concave Procedure	& QoS    	& Quality-of-Service \\
CIUSI  & Cascaded Inter-User- (IUI) \& Self-Interference (SI)	& RCG    & Riemannian Conjugate Gradient \\
CSI    & Channel State Information	& RB/TB    	& Reflect/Transmit Beamforming \\
DL/UL  & Downlink/Uplink				& RISI   	& Reverse ISI \\
EH     & Energy Harvesting		    & SCA    	& Successive Convex Approximation \\
FISI   & Forward Inter-System Interference (ISI)			& SDP    & Semi-Definite Program \\
GR     & Gaussian Randomization & SDR    	& Semi-Definite Relaxation \\
IET    & Interference Energy Threshold		& S(I)NR   & Signal-to-(Interference-plus-)Noise-Ratio \\
IRS    & Intelligent Reflecting Surface	& SR/ST & Secondary Receiver/Transmitter \\
JT(R)BPS  & Joint TB/(RB) \& Power Splitting (PS)	& SWIPT & Simultaneous Wireless Information \& Power Transfer \\	
\bottomrule
\end{tabular}}
\end{table}

\subsection{Structure, Mathematical Notation, and Abbreviations}\label{subsec:1.4}
The remainder of the paper is organized as follows: Sec.~\ref{sec:2} introduces the inter-system coordination protocol and the system model. Sec.~\ref{sec:3} presents the algorithms for the perfect CSI case, whereas Sec.~\ref{sec:4} focuses on outage-constrained robust design. Numerical simulation results are provided in Sec.~\ref{sec:5}. Our conclusions are discussed in Sec.~\ref{sec:6}.

\textbf{Notation:} $x$ or $X$: a scalar; $\mathbf{x}$: a column vector; $\mathbf{X}$: a matrix; $\mathbf{x}(n)$: the $n$-th element of $\mathbf{x}$; $\mathbf{X}(n,m)$: the $(n,m)$-th entry of $\mathbf{X}$; $\mathbb{R}^{N}$ and $\mathbb{C}^{N}$: the sets of real- and complex-valued $N$-dimensional vectors, respectively; $\mathbb{R}^{N\times M}$ and $\mathbb{C}^{N\times M}$: the sets of real- and complex-valued $N\times M$ matrices, respectively; $\mathbb{H}^{N}$: the set of Hermitian $N\times N$ matrices; $\left\|\mathbf{x}\right\|$: the Euclidean norm of $\mathbf{x}$; $\mathbf{X}^{*}$, $\mathbf{X}^{T}$, $\mathbf{X}^{\dagger}$, $\mathbf{X}^{-1}$, $\operatorname{Tr}\left(\mathbf{X}\right)$, $\operatorname{Rank}\left(\mathbf{X}\right)$, $\operatorname{vec}\left(\mathbf{X}\right)$, $\left\|\mathbf{X}\right\|_{2}$, and $\left\|\mathbf{X}\right\|_{*}$: the complex conjugate, transpose, complex conjugate transpose, inverse, trace, rank, vectorization, spectral norm, and nuclear norm of $\mathbf{X}$, respectively; $\operatorname{diag}\left(\mathbf{x}\right)$: a diagonal matrix whose main diagonal is $\mathbf{x}$; $\operatorname{Diag}\left(\mathbf{X}\right)$: a vector whose elements are extracted from the main diagonal entries of $\mathbf{X}$; $\mathbf{X}\succeq \mathbf{0}$: a positive semi-definite (PSD) matrix $\mathbf{X}$; $\mathbf{0}_{N}$: the $N$-dimensional null vector; $\mathbf{I}_{N}$: the $N\times N$ identity matrix; $\otimes$ and $\odot$: the Kronecker and Hadamard matrix product operators, respectively; $\lambda_{\max}\left(\mathbf{X}\right)$ and $\bm{\lambda}_{\max}\left(\mathbf{X}\right)$: the largest eigenvalue of $\mathbf{X}$ and the corresponding eigenvector; $j\triangleq\sqrt{-1}$: the imaginary unit; $\left|\cdot\right|$, $\operatorname{arg}\left(\cdot\right)$, and $\operatorname{Re}\left\{\cdot\right\}$: the magnitude, argument, and real part of a complex scalar, respectively; $\mathcal{CN}\left(\bm{\mu},\mathbf{C}\right)$: the circularly symmetric complex Gaussian (CSCG) distribution with mean $\bm{\mu}$ and covariance matrix $\mathbf{C}$; $\operatorname{Pr}(\cdot)$: the probability of an event; $\mathbb{E}\left\{\cdot\right\}$: the expectation operator; $\nabla_{\mathbf{x}}f\left(\mathbf{x}\right)$: the gradient vector with respect to (w.r.t.) $\mathbf{x}$; $\mathcal{O}\left(\cdot\right)$: the big-O notation.

\textbf{Abbreviations}: A list of the main abbreviations is given in Table~\ref{tab:tab1}.

\section{Coordination Protocol and System Model}\label{sec:2}

\subsection{System Setup}\label{subsec:2.1}
The secondary system consists of a ST with $M$ antennas, an IRS with $N$ passive reflecting elements, and $K$ active single-antenna SRs, while the primary system is comprised by a PT with $L$ antennas and $U$ active single-antenna PRs, as illustrated in Fig.~\ref{fig:1}. The sets of ST antennas, IRS elements, SRs, PT antennas, and PRs are denoted by $\mathcal{M}=\left\{1,\dots,M\right\}$, $\mathcal{N}=\left\{1,\dots,N\right\}$, $\mathcal{K}=\left\{1,\dots,K\right\}$, $\mathcal{L}=\left\{1,\dots,L\right\}$, and $\mathcal{U}=\left\{1,\dots,U\right\}$, respectively. Throughout the paper, we address the members of these sets via the indexes $m\in\mathcal{M}$, $n\in\mathcal{N}$, $k\in\mathcal{K}$, $l\in\mathcal{L}$, and $u\in\mathcal{U}$, unless it is explicitly indicated otherwise.

\subsection{Channel Model}\label{subsec:2.2}
We consider quasi-static, block-fading channels. The direct ST--SR $k$ and PT--PR $u$, reflect IRS--SR $k$ and IRS--PR $u$, and incident ST--IRS and PT--IRS baseband equivalent channels are denoted by $\mathbf{h}_{d,k}^{\dagger}\in\mathbb{C}^{M}$ and $\mathbf{g}_{d,u}^{\dagger}\in\mathbb{C}^{L}$, $\mathbf{h}_{r,k}^{\dagger}\in\mathbb{C}^{N}$ and $\mathbf{g}_{d,u}^{\dagger}\in\mathbb{C}^{N}$, and $\mathbf{H}\in\mathbb{C}^{N\times M}$ and $\mathbf{G}\in\mathbb{C}^{N\times L}$, respectively, while the direct ST--PR $u$ and PT--SR $k$ baseband equivalent channels are denoted as $\mathbf{v}_{d,u}^{\dagger}\in\mathbb{C}^{M}$ and $\mathbf{u}_{d,k}^{\dagger}\in\mathbb{C}^{L}$, respectively. 

We assume that the passive reflecting elements are arranged in a uniform rectangular array configuration as well as that the distance of any receiver from any transmitter or from the IRS is small, as it is typically the case in SWIPT and spectrum underlay setups or in IRS-aided wireless communication settings, respectively. Therefore, we adopt the spatially correlated Rician fading channel model. Thus, a direct transmitter-receiver or reflect IRS--receiver channel $\mathbf{c}$ and an incident transmitter--IRS channel $\mathbf{C}$ are described as $\mathbf{c} = \sqrt{C_{0}\left(\frac{d}{d_{0}}\right)^{-\alpha}}\left(\sqrt{\frac{\kappa}{\kappa +1}}\widebar{\mathbf{c}}+\sqrt{\frac{1}{\kappa +1}}\mathbf{R}^{1/2}\tilde{\mathbf{c}}\right)$ and $\mathbf{C} = \sqrt{C_{0}\left(\frac{D}{d_{0}}\right)^{-\beta}}\left(\sqrt{\frac{\varpi}{\varpi +1}}\widebar{\mathbf{C}}+\sqrt{\frac{1}{\varpi +1}}\mathbf{R}_{r}^{1/2}\widetilde{\mathbf{C}}\mathbf{R}_{t}^{1/2}\right)$, respectively, where $\left\{d,D\right\}$ is the separation distance of the corresponding nodes, $C_{0}$ denotes the free-space path loss at a reference distance $d_{0}=1$ m, $\left\{\alpha,\beta\right\}$ represents the respective path loss exponent, $\left\{\kappa,\varpi\right\}$ stands for the Rician factor, $\left\{\widebar{\mathbf{c}},\widebar{\mathbf{C}}\right\}$ refers to the deterministic line-of-sight (LoS) channel component, $\left\{\tilde{\mathbf{c}},\widetilde{\mathbf{C}}\right\}$ corresponds to the non-LoS (NLoS) channel component with independent and identically distributed (i.i.d.) $\mathcal{CN}(0,1)$ elements that capture the Rayleigh fading, $\left\{\mathbf{R}_{r},\mathbf{R}_{t}\right\}$ denote the receive and transmit correlation matrix of the incident channel, and $\mathbf{R}$ is the spatial correlation matrix of the considered reflect or direct channel.  

The RB matrix $\bm{\Theta}\in\mathbb{C}^{N\times N}$ and vector $\bm{\upsilon}\in\mathbb{C}^{N}$ are defined as $\bm{\Theta}\triangleq\operatorname{diag}\left(e^{j\theta_{1}},\dots,e^{j\theta_{N}}\right)$ and $\bm{\upsilon}\triangleq\operatorname{Diag}\left(\bm{\Theta}^{*}\right)$, respectively, where $\theta_{n}\in[0,2\pi)$ represents the phase shift of the $n$-th IRS element. The IRS-cascaded and effective ST--SR $k$ channels, $\mathbf{H}_{k}\in\mathbb{C}^{N\times M}$ and $\mathbf{h}_{k}^{\dagger}\in\mathbb{C}^{M}$, and their ST--PR $u$ counterparts, $\mathbf{V}_{u}\in\mathbb{C}^{N\times M}$ and $\mathbf{v}_{u}^{\dagger}\in\mathbb{C}^{M}$, are defined as $\mathbf{H}_{k}\triangleq\operatorname{diag}\left(\mathbf{h}_{r,k}^{\dagger}\right)\mathbf{H}$, $\mathbf{h}_{k}^{\dagger}\triangleq\mathbf{h}_{d,k}^{\dagger}+\mathbf{h}_{r,k}^{\dagger}\bm{\Theta}\mathbf{H}\triangleq\mathbf{h}_{d,k}^{\dagger}+\bm{\upsilon}^{\dagger}\mathbf{H}_{k}$, $\mathbf{V}_{u}\triangleq\operatorname{diag}\left(\mathbf{g}_{r,u}^{\dagger}\right)\mathbf{H}$, and $\mathbf{v}_{u}^{\dagger}\triangleq\mathbf{v}_{d,u}^{\dagger}+\mathbf{g}_{r,u}^{\dagger}\bm{\Theta}\mathbf{H}\triangleq\mathbf{v}_{d,u}^{\dagger}+\bm{\upsilon}^{\dagger}\mathbf{V}_{u}$, while IRS-cascaded and effective PT--PR $u$ channels, $\mathbf{G}_{u}\in\mathbb{C}^{N\times L}$ and $\mathbf{g}_{u}^{\dagger}\in\mathbb{C}^{L}$, and their PT--SR $k$ counterparts, $\mathbf{U}_{k}\in\mathbb{C}^{N\times L}$ and $\mathbf{u}_{k}^{\dagger}\in\mathbb{C}^{L}$, are defined by $\mathbf{G}_{u}\triangleq\operatorname{diag}\left(\mathbf{g}_{r,u}^{\dagger}\right)\mathbf{G}$, $\mathbf{U}_{k}\triangleq\operatorname{diag}\left(\mathbf{h}_{r,k}^{\dagger}\right)\mathbf{G}$, $\mathbf{g}_{u}^{\dagger}\triangleq\mathbf{g}_{d,u}^{\dagger}+\mathbf{g}_{r,u}^{\dagger}\bm{\Theta}\mathbf{G}\triangleq\mathbf{g}_{d,u}^{\dagger}+\bm{\upsilon}^{\dagger}\mathbf{G}_{u}$, and $\mathbf{u}_{k}^{\dagger}\triangleq\mathbf{u}_{d,k}^{\dagger}+\mathbf{h}_{r,k}^{\dagger}\bm{\Theta}\mathbf{G}\triangleq\mathbf{u}_{d,k}^{\dagger}+\bm{\upsilon}^{\dagger}\mathbf{U}_{k}$, $\forall k,u$, respectively.     

\subsection{Inter-System Coordination Protocol}\label{subsec:2.3}
The two systems operate over shared spectrum in time-division duplex (TDD) mode. The secondary system relies on joint TB/RB to compensate for the RISI and suppress the intra-system inter-user interference (IUI) at the SRs as well as to mitigate the FISI and the intra-primary-system ``IRS-cascaded'' inter-user- and self-interference (CIUSI) at the PRs. The PT, in turn, relies on TB to mitigate the intra-system IUI that reaches the PRs via direct PT--PR channels.

It is challenging to acquire CSI for the individual incident and reflect channels, due to the passive nature and limited processing capabilities of the IRS. Fortunately, estimating the instantaneous gains at the ST of the direct and IRS-cascaded DL channels of each transmitter with all receivers suffices for jointly optimizing the TB and RB weights~\cite{IRSCascCSI1,IRSCascCSI2}.

We assume an IRS hardware implementation where the passive reflecting elements can be in either OFF or ON state (i.e., full absorption or reflection), such that $\bm{\upsilon}(n)=0$ or $\bm{\upsilon}(n)=e^{j\theta_{n}}$, respectively~\cite{IRSSurvey,IRSChanEst}. Once set to ON state, we control only their phase shifts.

\begin{figure}[!t]
\centering
\captionsetup{justification=centering}
\includegraphics[scale = 0.5]{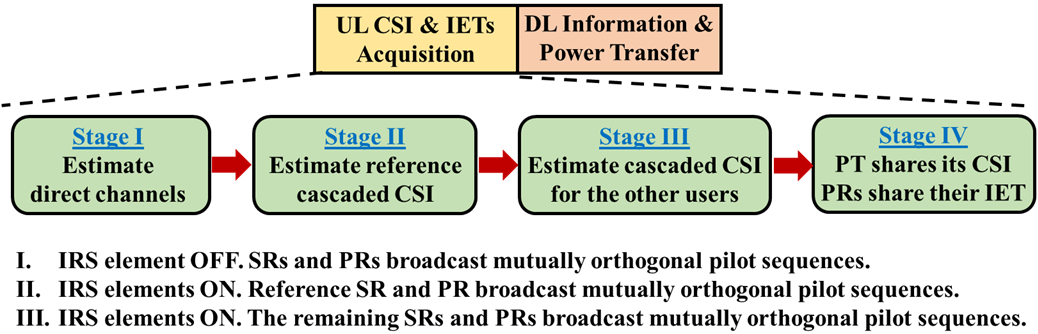}
\caption{Inter-system coordination protocol.}
\label{fig:Protocol}
\end{figure}

The transmission block is divided into an UL CSI (via training and channel estimation) and interference energy thresholds (IET) acquisition phase followed by a DL information and power transfer phase. We assume that the transmissions of the secondary and primary systems in the respective phases are perfectly synchronized. The applied CSI acquisition scheme corresponds to an adaptation of the three-stages method presented in~\cite{IRSChanEst}, in order to reduce the training overhead. Note that DL CSI acquisition is based on pilot-assisted estimation of the corresponding UL channels and exploitation of the channel reciprocity principle. The proposed coordinated side information acquisition protocol, which is shown in Fig.~\ref{fig:Protocol}, is described as follows: 1) All IRS elements are set to the OFF state and all SRs and PRs broadcast mutually orthogonal pilot sequences, such that the ST and the PT estimate their direct UL channels with all these users. 2) The ST selects a predefined RB vector and forwards it to both the PT and the IRS controller, which then sets all IRS elements to the ON state with the appropriate reflection coefficients. The ST and the PT also randomly select a reference SR and PR, respectively. These users broadcast mutually orthogonal pilot sequences, such that the ST and the PT estimate their effective UL channels with both and then infer the corresponding IRS-cascaded channels based on the knowledge of the reflections pattern and the respective direct channels acquired in stage (1). 3) The remaining SRs and PRs broadcast mutually orthogonal pilot sequences. The ST and PT efficiently estimate the respective IRS-cascaded UL channels with them by exploiting the fact that these channels are scaled versions of the corresponding reference CSI. 4) The PT and the PRs share its CSI and their IETs, respectively, with the ST.    

Based on the acquired side information, the ST jointly computes the optimal TB and RB weights and forwards the latter to the IRS controller, to enable SWIPT in the DL under the given objective function and constraints.   

The estimation of the direct and IRS-cascaded ST--SR, ST--PR, PT--PR, and PT--SR channels requires $\tau_{\mathbf{h}}=K+N+\left\lceil(K-1)N/M\right\rceil$, $\tau_{\mathbf{v}}=U+N+\left\lceil(U-1)N/M\right\rceil$, $\tau_{\mathbf{g}}=U+N+\left\lceil(U-1)N/L\right\rceil$, and $\tau_{\mathbf{u}}=K+N+\left\lceil(K-1)N/L\right\rceil$ symbols at minimum, respectively, where $\left\lceil\cdot\right\rceil$ denotes the ceiling operator~\cite{IRSChanEst}. Therefore, the minimum required pilot sequence length is given by $\tau_{p}=\max\left(\tau_{\mathbf{h}},\tau_{\mathbf{v}},\tau_{\mathbf{g}},\tau_{\mathbf{u}}\right)$. Assuming that $\tau_{s}$ slots are also used in stage IV for CSI and IETs sharing as well as that the transmission block has a length of $T$ symbols, it becomes apparent that a fraction $\tau/T$ of the transmission block is utilized for UL channel estimation purposes while the remaining fraction $\bar{\tau}\triangleq\left(T-\tau\right)/T$ is devoted to DL information and power transfer, where $\tau=\tau_{p}+\tau_{s}$. Under orthogonal transmissions in time, $\tau_{s}=K+U+1$.

\subsection{System Model}\label{subsec:2.4}

\subsubsection{Transmit Sum-Energy}\label{subsubsec:2.4.1}
The signals transmitted by the ST and the PT during the DL information and power transfer phase, $\mathbf{x}_{s}\in\mathbb{C}^{M}$ and $\mathbf{x}_{p}\in\mathbb{C}^{L}$, are given by $\mathbf{x}_{s} = \sum_{k\in\mathcal{K}}\mathbf{w}_{k}s_{k}$ and $\mathbf{x}_{p} = \sum_{u\in\mathcal{U}}\mathbf{f}_{u}c_{u}$, where $\mathbf{w}_{k}\in\mathbb{C}^{M}$ and $\mathbf{f}_{u}\in\mathbb{C}^{L}$ are the TB vectors assigned to SR $k$ and PR $u$, whereas $s_{k}\in\mathbb{C}$ which satisfies $\mathbb{E}\left\{\left|s_{k}\right|^{2}\right\} = 1$ and $c_{u}\sim\mathcal{CN}(0,1)$ denote the i.i.d. symbols transmitted to SR $k$ and PR $u$, respectively. Hence, the transmit sum-energy of the ST is given by $E_{s}=\bar{\tau}\sum_{k\in\mathcal{K}}\left\|\mathbf{w}_{k}\right\|^{2}$.  

\subsubsection{Achievable rate}\label{subsubsec:2.4.2}
\begin{figure}[!t]
\centering
\captionsetup{justification=centering}
\includegraphics[scale = 0.5]{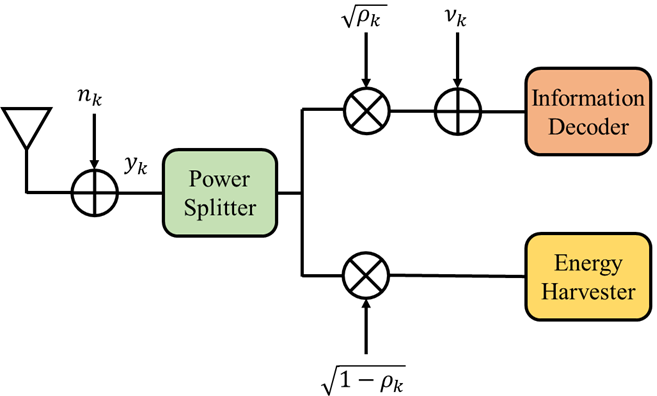}
\caption{PS-SWIPT receiver architecture of SR $k$.}
\label{fig:2}
\end{figure}

The PS-SWIPT receiver architecture is depicted in Fig.~\ref{fig:2}. We note that the RF power of the received signal is divided among two streams that are sent to the ID and EH units for data detection in the baseband and RF-to-DC energy conversion, respectively.

We assume availability of perfect CSI at the ST. The received signal at the ID branch of SR $k$ is given by $y_{k}^{I}=\sqrt{\rho_{k}}y_{k}+\nu_{k}=\sqrt{\rho_{k}}\left(\mathbf{h}_{k}^{\dagger}\mathbf{x}_{s}+\mathbf{u}_{d,k}^{\dagger}\mathbf{x}_{p}+n_{k}\right)+\nu_{k}$, where $0<\rho_{k}<1$ is the received PS ratio, $y_{k}$ denotes the received signal, $\nu_{k}\sim\mathcal{CN}\left(0,\sigma_{k,c}^{2}\right)$ stands for the baseband conversion noise at the ID branch, and $n_{k}\sim\mathcal{CN}\left(0,\sigma_{k}^{2}\right)$ represents the thermal noise. Hence, the SINR of SR $k$ is expressed as
\begin{equation}\label{eq:6}
\mathrm{SINR}_{k} = \frac{\left|\mathbf{h}_{k}^{\dagger}\mathbf{w}_{k}\right|^{2}}{\sum\limits_{i\in\mathcal{K}\setminus \{k\}}\left|\mathbf{h}_{k}^{\dagger}\mathbf{w}_{i}\right|^{2} + \hat{I}_{k} + \frac{\sigma_{k,c}^{2}}{\rho_{k}}}, \ \forall k,
\end{equation} 
where $\hat{I}_{k} \triangleq \tilde{I}_{k} + \sigma_{k}^{2}$ and $\tilde{I}_{k} \triangleq \sum_{u\in\mathcal{U}}\left|\mathbf{u}_{d,k}^{\dagger}\mathbf{f}_{u}\right|^{2}$. The achievable data rate of this user is given (in bps/Hz) by $R_{k}=\bar{\tau}\log_{2}\left(1+\mathrm{SINR}_{k}\right)$. 

\subsubsection{Harvested Energy}\label{subsubsec:2.4.3}
The received RF power at the input of the EH unit at SR $k$, omitting the negligible noise power, is expressed as $P_{k}^{\text{In}}=\left(1-\rho_{k}\right)P_{k}^{r} = \left(1-\rho_{k}\right)\left(\sum_{i\in\mathcal{K}}\left|\mathbf{h}_{k}^{\dagger}\mathbf{w}_{i}\right|^{2} + \tilde{I}_{k}\right)$, where $P_{k}^{r}$ stands for the received RF power at SR $k$. Therefore, the received RF energy at the EH branch is equal to $E_{k}^{\text{In}}=\bar{\tau}P_{k}^{\text{In}}$. The harvested DC energy at the output the EH unit, in turn, is written as $E_{k}^{\text{Out}}=F\left(E_{k}^{\text{In}}\right)$, where $F\left(\cdot\right)$ denotes a monotonically increasing EH function that captures the non-linearities of the diodes-based rectifier circuit. Specifically, we adopt the model $F(x)=\left(\bar{a}x+\bar{b}\right)/\left(x+\bar{c}\right)-\bar{b}/\bar{c}$, where the harsware-dependent parameters $\bar{a}$, $\bar{b}$, $\bar{c}$ capture the saturation and sensitivity thresholds of the EH circuit as well as other properties, such as its capacitance and resistance, and they can be estimated by standard curve fitting algorithms~\cite{SAR}.

\section{Proposed Designs Under Perfect CSI}\label{sec:3}

\subsection{Problem Formulation}\label{subsec:3.1}
We aim at performing JTRBPS optimization such that the transmit sum-energy is minimized s.t. the rate or, equivalently, SINR and EH constraints of the SRs, the FISI and CIUSI interference constraints of the PRs, the constraints of the receive PS ratios, and the non-convex unit modulus constraints of the RB weights. These constraints are described as:
\begin{subequations}\label{eq:reqs}
\begin{alignat}{2}
&R_{k} \geq R_{\min,k} \Leftrightarrow \text{C1: }\mathrm{SINR}_{k} \geq 2^{\frac{R_{\min,k}}{\bar{\tau}}}-1\triangleq\Gamma_{k},  \ \forall k, \label{eq:reqsa} \\
&E_{k}^{\text{Out}}\geq\tilde{Q}_{k} \Leftrightarrow \text{C2: }E_{k}^{\text{In}}\geq F^{-1}\left(\tilde{Q}\right)\triangleq Q_{k}, \ \forall k, \label{eq:reqsb} \\
&\text{C3: }\mathrm{FISI}_{u} = \bar{\tau}\sum_{k\in\mathcal{K}}\left|\mathbf{v}_{d,u}^{\dagger}\mathbf{w}_{k}\right|^{2} \leq E_{u}^{I}, \ \forall u, \label{eq:reqsc} \\
&\text{C4: }\mathrm{CIUSI}_{u} = \bar{\tau}\sum_{j\in\mathcal{U}}\left|\bm{\upsilon}^{\dagger}\mathbf{G}_{u}\mathbf{f}_{j}\right|^{2} \leq E_{\text{Th}}, \ \forall u, \label{eq:reqsd} \\
& \text{C5: } 0\leq\rho_{k}\leq 1, \ \forall k, \ \text{C6: } \left|\bm{\upsilon}(n)\right|=1, \ \forall n, \label{eq:reqse}
\end{alignat}
\end{subequations}
where $R_{\min,k}>0$ and $\tilde{Q}_{k}>0$ are the minimum rate and DC harvested energy thresholds of SR $k$ while $\Gamma_{k}>0$ and $Q_{k}>0$ denote the corresponding minimum SINR and RF power at the input of the EH unit thresholds, respectively, and $E_{u}^{I}>0$ and $E_{\text{Th}}$ represent the IET and the (arbitrarily small) CIUSI energy threshold of PR $u$, respectively. Hence, the JTRBPS optimization problem is formulated accordingly:
\begin{equation}\label{eq:10}
\text{(P1):}\ \!\underset{\left\{\mathbf{w}_{k}\right\},\bm{\upsilon},\left\{\rho_{k}\right\}}{\min} \ E_{s} \ \text{s.t. C1--C6}.
\end{equation}
(P1) is a challenging non-convex optimization problem, due to the intrinsic coupling of the optimization variables in the constraints C1--C4 and the non-convex constraints C6. In order to tackle it, we develop two iterative algorithms.

\subsection{Penalty-Based Alternating Minimization Algorithm}\label{subsec:3.2}
In the first algorithm, we apply the AM principle to decouple the optimization variables, i.e., we alternate between updating the TB weights and PS ratios for fixed RB weights and vice-versa in an iterative manner until convergence or the maximum number of iterations is reached. 

\subsubsection{JTBPS Optimization}\label{subsubsec:3.2.1}
For fixed $\bm{\upsilon}$, we have:
\begin{equation}\label{eq:11}
\text{(P2):}\ \!\underset{\left\{\mathbf{w}_{k}\right\},\left\{\rho_{k}\right\}}{\min} \ E_{s} \ \text{s.t. \ C1--C3, C5.}
\end{equation} 

We apply SDR to solve (P2), i.e., we introduce rank-one PSD matrix variables $\mathbf{W}_{k}\in\mathbb{H}^{M}$ defined as $\mathbf{W}_{k} \triangleq \mathbf{w}_{k}\mathbf{w}_{k}^{\dagger}$ and then we drop the non-convex rank constraints to obtain:  
\begin{subequations}\label{eq:12}
\begin{alignat}{2}
&&& \text{(P3):}\ \!\underset{\left\{\mathbf{W}_{k}\right\},\left\{\rho_{k}\right\}}{\min} \ \bar{\tau}\sum_{k\in\mathcal{K}}\operatorname{Tr}\left(\mathbf{W}_{k}\right) \label{eq:12a} \\
&&&\text{s.t.} \ \ \ \overline{\text{C1}}\text{: }\frac{\operatorname{Tr}\left(\mathbf{h}_{k}\mathbf{h}_{k}^{\dagger}\mathbf{W}_{k}\right)}{\sum\limits_{i\in\mathcal{K}\setminus \{k\}}\operatorname{Tr}\left(\mathbf{h}_{k}\mathbf{h}_{k}^{\dagger}\mathbf{W}_{i}\right) + I_{k}} \geq \frac{\Gamma_{k}}{1+\Gamma_{k}}, \ \forall k, \label{eq:12b} \\
&&& \ \ \ \ \ \ \ \overline{\text{C2}}\text{: }\sum_{i\in\mathcal{K}}\operatorname{Tr}\left(\mathbf{h}_{k}\mathbf{h}_{k}^{\dagger}\mathbf{W}_{i}\right) + \tilde{I}_{k} \geq \frac{Q_{k}}{\bar{\tau}\left(1-\rho_{k}\right)}, \ \forall k, \label{eq:12c} \\
&&& \ \ \ \ \ \ \ \overline{\text{C3}}\text{: }\sum_{k\in\mathcal{K}}\operatorname{Tr}\left(\mathbf{v}_{d,u}\mathbf{v}_{d,u}^{\dagger}\mathbf{W}_{k}\right) \leq \frac{E_{u}^{I}}{\bar{\tau}}, \ \forall u, \label{eq:12d} \\
&&& \ \ \ \ \ \ \ \text{C5: }0 \leq \rho_{k} \leq 1, \ \forall k, \ \text{C7: }\mathbf{W}_{k} \succeq \mathbf{0}, \ \forall k, \ \text{C8: }\cancelto{}{\operatorname{Rank}\left(\mathbf{W}_{k}\right)=1}, \ \forall k, \label{eq:12e}
\end{alignat}
\end{subequations}
where $I_{k} \triangleq \hat{I}_{k} + \sigma_{k,c}^{2}/\rho_{k}$. (P3) is a convex semi-definite program (SDP). Thus, it can be solved in polynomial time by convex optimization software, such as CVX~\cite{CVX}. By following the approach in~\cite{Zhang2014a}, it is trivial to show that $\operatorname{Rank}\left(\mathbf{W}_{k}^{\star}\right) = 1$, $\forall k$, such that we can extract $\mathbf{w}_{k}^{\star}$ via eigen-value decomposition.

\subsubsection{RB Optimization}\label{subsubsec:3.2.2}
For given $\left\{\mathbf{w}_{k},\rho_{k}\right\}$, the optimization of $\bm{\upsilon}$ is reduced to a feasibility check problem:
\begin{equation}\label{eq:13}
\text{(P4):}\ \!\text{Find } \bm{\upsilon} \ \text{s.t. C1--C4, C6.}
\end{equation} 

Let $\bar{\bm{\upsilon}}\triangleq\left[\bm{\upsilon};x\right]$, where $\widebar{N}\triangleq N+1$ and $x$ with $\left|x\right|^{2}=1$ is an auxiliary variable, and $k,i\in\mathcal{K}$, $u,j\in\mathcal{U}$. Also, let $a_{k,i}\triangleq\mathbf{h}_{d,k}^{\dagger}\mathbf{w}_{i}$ and $b_{u,i}\triangleq\mathbf{v}_{d,u}^{\dagger}\mathbf{w}_{i}$. Furthermore, let us define $\left\{\mathbf{a}_{k,i},\mathbf{b}_{u,i},\mathbf{c}_{u,j}\right\}\in\mathbb{C}^{N}$ and $\left\{\bm{\Phi}_{k,i},\bm{\Psi}_{u,i},\bm{\Omega}_{u,j}\right\}\in\mathbb{C}^{\widebar{N}\times\widebar{N}}$ as $\mathbf{a}_{k,i}\triangleq\mathbf{H}_{k}\mathbf{w}_{i}, \ \mathbf{b}_{u,i}\triangleq\mathbf{V}_{u}\mathbf{w}_{i}, \ \mathbf{c}_{u,j}\triangleq\mathbf{G}_{u}\mathbf{f}_{j}$ and $\bm{\Phi}_{k,i}\triangleq\left[\mathbf{a}_{k,i}\mathbf{a}_{k,i}^{\dagger}, \mathbf{a}_{k,i}a_{k,i}^{*}; \mathbf{a}_{k,i}^{\dagger}a_{k,i},0\right]$, $\bm{\Psi}_{u,i}\triangleq\left[\mathbf{b}_{u,i}\mathbf{b}_{u,i}^{\dagger}, \mathbf{b}_{u,i}b_{u,i}^{*}; \mathbf{b}_{u,i}^{\dagger}b_{u,i},0\right]$, and $\bm{\Omega}_{u,j}\triangleq\left[\mathbf{c}_{u,j}\mathbf{c}_{u,j}^{\dagger}, \mathbf{0}_{N}; \mathbf{0}_{N}^{\dagger},0\right]$, respectively. Next, we introduce the rank-one PSD matrix variable $\mathbf{V}\in\mathbb{H}^{\widebar{N}}$ defined as $\mathbf{V} \triangleq \bar{\bm{\upsilon}}\bar{\bm{\upsilon}}^{\dagger}$ and convert (P4) to:
\begin{subequations}\label{eq:extra}
\begin{alignat}{2}
&&& \text{(P5):}\ \!\text{Find } \mathbf{V} \label{eq:extraa} \\
&&&\text{s.t.} \ \ \ \overline{\overline{\text{C1}}}\text{: } \operatorname{Tr}\left(\bm{\Phi}_{k,k}\mathbf{V}\right) + \left|a_{k,k}\right|^{2} \geq \Gamma_{k}\left(\sum\limits_{i\in\mathcal{K}\setminus \{k\}}\left(\operatorname{Tr}\left(\bm{\Phi}_{k,i}\mathbf{V}\right) + \left|a_{k,i}\right|^{2}\right) + I_{k}\right), \ \forall k, \label{eq:extrab} \\
&&& \ \ \ \ \ \ \ \overline{\overline{\text{C2}}}\text{: } \sum\limits_{i\in\mathcal{K}}\left(\operatorname{Tr}\left(\bm{\Phi}_{k,i}\mathbf{V}\right) + \left|a_{k,i}\right|^{2}\right) + \tilde{I}_{k} \geq \frac{\widebar{Q}_{k}}{\left(1-\rho_{k}\right)}, \ \forall k, \label{eq:extrac} \\ 
&&& \ \ \ \ \ \ \ \overline{\overline{\text{C3}}}\text{: } \sum_{k\in\mathcal{K}}\left(\operatorname{Tr}\left(\bm{\Psi}_{u,k}\mathbf{V}\right) + \left|b_{u,k}\right|^{2}\right) \leq \widebar{E}_{u}^{I}, \ \forall u, \label{eq:extrad} \\
&&& \ \ \ \ \ \ \ \overline{\text{C4}}\text{: }\sum_{j\in\mathcal{U}}\operatorname{Tr}\left(\bm{\Omega}_{u,j}\mathbf{V}\right) \leq \widebar{E}_{\text{Th}}, \ \forall u, \label{eq:extrae} \\
&&& \ \ \ \ \ \ \ \text{C9: }\mathbf{V}(n,n) = 1, \ n=1,\dots,\widebar{N}, \ \text{C10: } \mathbf{V} \succeq \mathbf{0}, \ \text{C11: }\operatorname{Rank}\left(\mathbf{V}\right)=1, \label{eq:extraf}
\end{alignat}
\end{subequations}
where we have defined $\widebar{Q}_{k}\triangleq Q_{k}/\bar{\tau}$, $\widebar{E}_{u}^{I}\triangleq E_{u}^{I}/\bar{\tau}$, and $\widebar{E}_{\text{Th}}\triangleq E_{\text{Th}}/\bar{\tau}$ for convenience. The standard approach for tackling problem (P5) in the literature is to employ the SDR method, i.e., to relax the non-convex rank-one constraint C11, such that (P5) is transformed into a convex SDP which can be efficiently solved by CVX to obtain $\mathbf{V}^{\star}$. Generally, though, $\operatorname{Rank}\left(\mathbf{V}^{\star}\right)\neq 1$. Thus, the GR method is commonly used to construct a sub-optimal rank-one solution $\bar{\bm{\upsilon}}^{\star}$. Then, we can recover the corresponding solution of the original problem (P4), $\bm{\upsilon}^{\star}$, as $\bm{\upsilon}^{\star}=e^{j\operatorname{arg}\left(\bar{\bm{\upsilon}}_{1:N}/\bar{\bm{\upsilon}}\left(\widebar{N}\right)\right)}$, where $\mathbf{x}_{1:N}$ refers to the first $N$ elements of vector $\mathbf{x}$. This SDR approach followed by a sufficiently large number of GR rounds guarantees a $\pi/4$-approximation of the optimal objective value of the original problem (P4)~\cite{GaussianRand}.

Nevertheless, this strategy has two major drawbacks~\cite{Provable,Provable2}: i) by solving the feasibility check problem (P5), we cannot guarantee convergence of the AM algorithm; and ii) GR does not necessarily produce a feasible solution (i.e., one that meets the constraints). Hence, we follow a different approach to avoid both the feasibility check problem formulation (P5) and the adoption of the SDR/GR mechanism~\cite{Provable,Provable2}. Specifically, we equivalently express the rank-one constraint C11 as $\overline{\text{C11}}\text{: }\left\|\mathbf{V}\right\|_{*}-\left\|\mathbf{V}\right\|_{2} \leq 0$, where $\left\|\mathbf{V}\right\|_{*} = \sum_{i}\sigma_{i}\left(\mathbf{V}\right)\geq\left\|\mathbf{V}\right\|_{2}=\max_{i}\left\{\sigma_{i}\left(\mathbf{V}\right)\right\}$ and $\sigma_{i}\left(\mathbf{V}\right)$ denotes the $i$-th singular value of $\mathbf{V}$. The equality holds if and only if $\mathbf{V}$ is rank-one. Therefore, we integrate next a penalized version of $\overline{\text{C11}}$ into the objective function of problem (P5). The resulting optimization problem has the following form:
\begin{equation}\label{eq:extra2}
\text{(P6):}\ \!\underset{\mathbf{V}}{\min} \frac{1}{2\mu}\left(\left\|\mathbf{V}\right\|_{*}-\left\|\mathbf{V}\right\|_{2}\right) \ \text{s.t.} \ \overline{\overline{\text{C1}}}\text{--}\overline{\overline{\text{C3}}}, \ \overline{\text{C4}}, \text{C9, C10},
\end{equation}
where $\mu>0$ is a factor that penalizes the violations of C11. Notice that although we have relaxed the rank-one constraint C11 in problem (P6), the solution obtained for $\mu\rightarrow 0$ is an optimal solution of (P5) as well~\cite{Provable,Provable2}. At the same time, for sufficiently small penalty factor $\mu$, the solution of (P6) is of rank one~\cite{Provable,Provable2}.

However, (P6) is a non-convex optimization problem, since the objective function consists of the difference of convex (D.C.) functions. We can address this issue by applying the SCA technique. In particular, in the $(t+1)$-th iteration of the SCA algorithm, we construct a global underestimator of the objective function by leveraging the first-order Taylor approximation $\left\|\mathbf{V}\right\|_{2}\geq\left\|\widebar{\mathbf{V}}^{(t)}\right\|_{2}+\operatorname{Tr}\left(\bm{\lambda}_{\max}\left(\widebar{\mathbf{V}}^{(t)}\right)\bm{\lambda}_{\max}^{\dagger}\left(\widebar{\mathbf{V}}^{(t)}\right)\left(\mathbf{V}-\widebar{\mathbf{V}}^{(t)}\right)\right)$ $\triangleq\widebar{V}^{(t+1)}$, where $\widebar{\mathbf{V}}^{(t)}$ denotes the solution obtained in the $t$-th iteration. Therefore, we recast (P6) in the $(t+1)$-th iteration as:
\begin{equation}\label{eq:extra3}
\text{(P7):}\ \!\underset{\mathbf{V}}{\min} \frac{1}{2\mu}\left(\left\|\mathbf{V}\right\|_{*}-\widebar{V}^{(t+1)}\right) \ \text{s.t.} \ \overline{\overline{\text{C1}}}\text{--}\overline{\overline{\text{C3}}}, \ \overline{\text{C4}}, \text{C9, C10}.
\end{equation}   
The implicit feasibility check problem (P7) is a convex SDP and can be efficiently solved by CVX. Then, we can recover $\bar{\bm{\upsilon}}^{\star}$ (and, therefore, $\bm{\upsilon}^{\star}$, as described earlier) from the Cholesky decomposition of the rank-one matrix $\mathbf{V}^{\star}=\bar{\bm{\upsilon}}^{\star}\left(\bar{\bm{\upsilon}}^{\star}\right)^{\dagger}$.

\subsubsection{Convergence and Computational Complexity}\label{subsubsec:3.2.3}
The SCA algorithm associated with problem (P7) is presented in Alg.~\ref{algo:algoSCA}, while the overall penalty-based AM algorithm for solving problem (P1) is described in Alg.~\ref{algo:algo1}. Note that the proposed AM algorithm provably generates a non-increasing sequence of objective values and converges to a stationary point of problem (P1) in polynomial time~\cite{Provable,Provable2}. The interior-point method algorithm used to obtain an $\epsilon$-optimal solution of the SDP (P3) requires at most $\mathcal{O}\left(\sqrt{n}\log(1/\epsilon)\right)$ iterations, where $n=M$~\cite{IPM}. Similarly, for solving problem (P7) are required at most $\mathcal{O}\left(\sqrt{n}I_{\text{SCA}}\log(1/\epsilon)\right)$ iterations, where $n=\widebar{N}$ and $I_{\text{SCA}}$ denotes the number of iterations of the SCA algorithm~\cite{IPM}. The computation cost per iteration is given in both cases by $\mathcal{O}\left(mn^{3}+m^{2}n^{2}+m^{3}\right)$, where $m=3K+U$ in (P3) and $m=2K+2U+1$ in  (P7)~\cite{IPM}.
\begin{algorithm}[!t]
\centering \footnotesize
  \begin{algorithmic}[1]
  \State Initialize $\widebar{\mathbf{V}}^{(0)}$ with $\widebar{N}$ random phase shifts. Set the convergence tolerance $0 \leq \epsilon \ll 1$, penalty factor $0 \leq \mu \ll 1$, iteration index $t=0$, and maximum number of iterations $T_{\max}$;
  \Repeat
  	\State Solve problem (P7) with given $\widebar{\mathbf{V}}^{(t)}$ to obtain $\widebar{\mathbf{V}}^{(t+1)}$;
  	\State $t \leftarrow t + 1$;
  \Until
  	\State $\widebar{N}-\lambda_{\max}\left(\mathbf{V}^{(t)}\right)/\widebar{N} \leq \epsilon$ or $t=T_{\max}$;
  	\State Output: $\widebar{\mathbf{V}}^{\star}$;
\end{algorithmic}
\caption{SCA Algorithm for Solving Problem (P7).}
\label{algo:algoSCA}
\end{algorithm}
\begin{algorithm}[!t]
\centering \footnotesize
  \begin{algorithmic}[1]
  \State Initialize $\mathbf{V}^{(0)}$ with $\widebar{N}$ random phase shifts. Set the convergence tolerance $0 \leq \epsilon \ll 1$, iteration index $j=0$, and maximum number of iterations $J_{\max}$;
  \Repeat
  	\State Solve problem (P3) with given $\mathbf{V}^{(j)}$ to obtain $\left\{\mathbf{w}_{k}^{(j+1)}, \rho_{k}^{(j+1)}\right\}$;
  	\State Apply Alg.~\ref{algo:algoSCA} with given $\left\{\mathbf{w}_{k}^{(j+1)}, \rho_{k}^{(j+1)}\right\}$ to obtain $\mathbf{V}^{(j+1)}=\widebar{\mathbf{V}}^{\star}$ 
  	\State $j = j + 1$;
  \Until
  	\State $\left|f_{j+1}-f_{j}\right|/\left|f_{j}\right| \leq \epsilon $, where $f_{j} = \bar{\tau}\sum\limits_{k\in\mathcal{K}}\left\|\mathbf{w}_{k}^{(j)}\right\|^{2}$, or $j=J_{\max}$;
   \State Output: $\left\{\mathbf{w}_{k}^{\star},\rho_{k}^{\star},\mathbf{V}^{\star}\right\}$. 
\end{algorithmic}
\caption{AM Algorithm for Solving Problem (P1).}
\label{algo:algo1}
\end{algorithm}

\subsection{Penalty-Based BCD Algorithm}\label{subsec:3.3}
In the second algorithm, we reformulate the problem to decouple the optimization variables and then we apply the BCD framework to solve the resulting optimization problem.

\subsubsection{Problem Reformulation}\label{subsubsec:3.3.1}
Let $\mathbf{h}_{k}^{\dagger}\mathbf{w}_{i}=t_{k,i}$, $\mathbf{v}_{d,u}^{\dagger}\mathbf{w}_{k}=\psi_{u,k}$, and $\bm{\upsilon}^{\dagger}\mathbf{G}_{u}\mathbf{f}_{j} = \lambda_{u,j}$ $\forall k,i\in\mathcal{K}$, $\forall u,j\in\mathcal{U}$. Then, problem (P1) is converted to:
\begin{subequations}\label{eq:BCD1}
\begin{alignat}{2}
&&&\text{(P8):}\ \!\underset{\left\{\mathbf{w}_{k}\right\},\bm{\upsilon},\left\{\rho_{k}\right\}}{\min} \ \bar{\tau}\sum_{k\in\mathcal{K}}\left\|\mathbf{w}_{k}\right\|^{2} \label{eq:BCD1a} \\
&&&\text{s.t.} \ \ \ \text{C12: } \frac{\left|t_{k,k}\right|^{2}}{\sum\limits_{i\in\mathcal{K}\setminus\{k\}}\left|t_{k,i}\right|^{2} + I_{k}} \geq \Gamma_{k}, \ \forall k, \label{eq:BCD1b} \\
&&& \ \ \ \ \ \ \ \text{C13: } \sum_{i\in\mathcal{K}}\left|t_{k,i}\right|^{2} + \tilde{I}_{k} \geq \frac{Q_{k}}{\bar{\tau}\left(1-\rho_{k}\right)}, \ \forall k, \label{eq:BCD1c} \\
&&& \ \ \ \ \ \ \ \text{C14: } \sum_{k\in\mathcal{K}}\left|\psi_{u,k}\right|^{2} \leq \frac{E_{u}^{I}}{\bar{\tau}}, \ \forall u, \label{eq:BCD1d} \\
&&& \ \ \ \ \ \ \ \text{C15: } \sum_{j\in\mathcal{U}}\left|\lambda_{u,j}\right|^{2} \leq \frac{E_{\text{Th}}}{\bar{\tau}}, \ \forall u, \label{eq:BCD1e} \\
&&& \ \ \ \ \ \ \ \text{C16: } \mathbf{h}_{k}^{\dagger}\mathbf{w}_{i}=t_{k,i}, \ \forall k,i, \ \text{C17: } \mathbf{v}_{d,u}^{\dagger}\mathbf{w}_{k}=\psi_{u,k}, \ \forall u,k, \ \text{C18: } \bm{\upsilon}^{\dagger}\mathbf{G}_{u}\mathbf{f}_{j} = \lambda_{u,j}, \ \forall u,j, \label{eq:BCD1f} \\
&&& \ \ \ \ \ \ \ \text{C19: } 0\leq\rho_{k}\leq 1, \ \forall k, \text{C20: } \left|\bm{\upsilon}(n)\right|=1, \ \forall n. \label{eq:BCD1i}
\end{alignat}
\end{subequations}
Converting the equality constraints C16--C18 to quadratic functions and adding them as penalty terms to the objective function of problem (P8) yields:
\begin{subequations}\label{eq:BCD2}
\begin{alignat}{2}
&&&\text{(P9):}\ \!\underset{\substack{\left\{\mathbf{w}_{k}\right\},\bm{\upsilon},\left\{\rho_{k}\right\},\\ \left\{t_{k,i}\right\},\left\{\psi_{u,k}\right\},\left\{\lambda_{u,j}\right\}}}{\min} \ \bar{\tau}\sum_{k\in\mathcal{K}}\left\|\mathbf{w}_{k}\right\|^{2} + \frac{1}{2\omega}\left(\sum_{k\in\mathcal{K}}\sum_{i\in\mathcal{K}}\left|\mathbf{h}_{k}^{\dagger}\mathbf{w}_{i}-t_{k,i}\right|^{2} +\sum_{u\in\mathcal{U}}\sum_{k\in\mathcal{K}}\left|\mathbf{v}_{d,u}^{\dagger}\mathbf{w}_{k}-\psi_{u,k}\right|^{2}
\right. \nonumber \\
&&& \quad \quad \quad \quad \quad \quad \quad \quad \quad  \left. +\sum_{u\in\mathcal{U}}\sum_{j\in\mathcal{U}}\left|\bm{\upsilon}^{\dagger}\mathbf{G}_{u}\mathbf{f}_{j}-\lambda_{u,j}\right|^{2}\right) \label{eq:BCD2a} \\
&&&\text{s.t.} \ \ \ \text{C12--C15, C19, C20}, \label{eq:BCD2b}
\end{alignat}
\end{subequations}
where $\omega>0$ is a factor that penalizes the violations of the equality constraints C16--C18. We notice that we can partition the optimization variables into distinct blocks and apply the BCD method to efficiently solve this non-convex optimization problem iteratively by alternately optimizing each block in one iteration with the other blocks fixed until convergence or the maximum number of iterations is reached.

\subsubsection{TB Optimization}\label{subsubsec:3.3.2}
With all blocks except $\left\{\mathbf{w}_{k}\right\}$ fixed, problem (P9) is reduced (by ignoring the constant terms) to:
\begin{align}\label{eq:BCD3}
&\text{(P10):}\ \!\underset{\left\{\mathbf{w}_{k}\right\}}{\min} \ \bar{\tau}\sum_{k\in\mathcal{K}}\left\|\mathbf{w}_{k}\right\|^{2} + \frac{1}{2\omega}\left(\sum_{k\in\mathcal{K}}\sum_{i\in\mathcal{K}}\left|\mathbf{h}_{k}^{\dagger}\mathbf{w}_{i}-t_{k,i}\right|^{2} +\sum_{u\in\mathcal{U}}\sum_{k\in\mathcal{K}}\left|\mathbf{v}_{d,u}^{\dagger}\mathbf{w}_{k}-\psi_{u,k}\right|^{2}\right)
\end{align}
(P10) is a convex unconstrained quadratic minimization problem. Based on the first-order optimallity condition, we obtain the solution in closed-form expression as:
\begin{equation}\label{eq:BCD4}
\mathbf{w}_{k}^{\star} = \frac{1}{2\omega}\mathbf{A}^{-1}\left(\sum_{i\in\mathcal{K}}\mathbf{h}_{i}^{\dagger}t_{i,k}+\sum_{u\in\mathcal{U}}\mathbf{v}_{d,u}^{\dagger}\psi_{u,k}\right), \ \forall k,
\end{equation} 
where $\mathbf{A}\in\mathbb{C}^{M\times M}$ is given by $\mathbf{A}=\mathbf{I}_{M}+\frac{1}{2\omega}\sum_{i\in\mathcal{K}}\mathbf{h}_{i}\mathbf{h}_{i}^{\dagger} + \sum_{u\in\mathcal{U}}\mathbf{v}_{d,u}\mathbf{v}_{d,u}^{\dagger}$.

\subsubsection{RB Optimization}\label{subsubsec:3.3.3}
Let us define $a_{k,i}$, $\mathbf{a}_{k,i}\in\mathbf{C}^{N}$, and $\mathbf{c}_{u,j}\in\mathbb{C}^{N}$ as $a_{k,i}\triangleq\mathbf{h}_{d,k}^{\dagger}\mathbf{w}_{i}$, $\mathbf{a}_{k,i}\triangleq\mathbf{H}_{k}\mathbf{w}_{i}$, and $\mathbf{c}_{u,j}=\mathbf{G}_{u}\mathbf{f}_{j}$, $\forall k,i\in\mathcal{K}$, $\forall u,j\in\mathcal{U}$, as in Sec.~\ref{subsubsec:3.2.2}. Then, with all blocks except $\bm{\upsilon}$ fixed, problem (P9) is reduced (by ignoring the constant terms) to
\begin{equation}\label{eq:BCD5}
\text{(P11):}\ \!\underset{\bm{\upsilon}}{\min} \ f\left(\bm{\upsilon}\right)=\sum_{k\in\mathcal{K}}\sum_{i\in\mathcal{K}}\left|\bm{\upsilon}^{\dagger}\mathbf{a}_{k,i}+a_{k,i}-t_{k,i}\right|^{2} +\sum_{u\in\mathcal{U}}\sum_{j\in\mathcal{U}}\left|\bm{\upsilon}^{\dagger}\mathbf{c}_{u,j}-\lambda_{u,j}\right|^{2} \text{ s.t. } \text{C20}. 
\end{equation}
Note  that the Euclidean gradient of $f\left(\bm{\upsilon}\right)$ over $\bm{\upsilon}$ is given by
\begin{equation}\label{eq:BCD6}
\nabla f\left(\bm{\upsilon}\right) = 2\sum_{k\in\mathcal{K}}\sum_{i\in\mathcal{K}}\mathbf{a}_{k,i}\left(\mathbf{a}_{k,i}\bm{\upsilon}+a_{k,i}^{*}-t_{k,i}^{*}\right) + 2\sum_{u\in\mathcal{U}}\sum_{j\in\mathcal{U}}\mathbf{c}_{u,j}\left(\mathbf{c}_{u,j}^{\dagger}\bm{\upsilon}-\lambda_{u,j}^{*}\right).
\end{equation} 
The unit-modulus constraints C20 in problem (P11) form a Riemannian manifold: 
\begin{equation*}
\mathcal{S}^{N}: \left\{\bm{\upsilon}\in\mathbb{C}^{N}: \left|\bm{\upsilon}(n)\right|=1\right\}.
\end{equation*}
Hence, (P11) can be viewed as a minimization problem over the considered search space and solved iteratively by manifold optimization methods via the following steps at each iteration: i) We compute the Riemannian gradient as the tangent vector $\mathbf{g}$ given by the orthogonal projection of $\nabla f\left(\bm{\upsilon}\right)$ onto the tangent space: $\mathbf{g}=\nabla f\left(\bm{\upsilon}\right)-\operatorname{Re}\left\{\nabla f\left(\bm{\upsilon}\right)\odot\bm{\upsilon}^{*}\right\}\odot\bm{\upsilon}$. ii) By using the RCG algorithm, we update the search direction as $\mathbf{d} = - \mathbf{g} + \gamma\mathcal{T}\left(\bar{\mathbf{d}}\right)$, where $\mathcal{T}\left(\mathbf{d}\right)\triangleq\bar{\mathbf{d}}-\operatorname{Re}\left\{\mathbf{d}\odot\bm{\upsilon}^{*}\right\}\odot\bm{\upsilon}$ denotes the vector transport function, $\gamma$ is the Polak-Ribiere conjugate gradient update parameter, and $\bar{\mathbf{d}}$ stands for the previous search direction. iii) Finally, we update the RB weights via retraction as $\bm{\upsilon} \leftarrow \frac{\left(\bm{\upsilon}+\eta\mathbf{d}\right)_{n}}{\left|\left(\bm{\upsilon}+\eta\mathbf{d}\right)_{n}\right|}$, where $\eta$ represents the step size of Armijo backtracking line search~\cite{ManOptBook}.

\subsubsection{PS Ratios and Auxiliary Variables Optimization}\label{subsubsec:3.3.4}
With all variables fixed except $\left\{\rho_{k}\right\}$ and $\left\{t_{k,i}\right\}$, problem (P9) is reduced (by ignoring the constant terms) to:
\begin{equation}\label{eq:BCD7}
\text{(P12):}\ \!\underset{\left\{\rho_{k}\right\},\left\{t_{k,i}\right\}}{\min}\ \sum_{k\in\mathcal{K}}\sum_{i\in\mathcal{K}}\left|\mathbf{h}_{k}^{\dagger}\mathbf{w}_{i}-t_{k,i}\right|^{2} \text{ s.t. C12, C13, C19}. 
\end{equation}
We initially rearrange the terms in constraint C12, in order to solve problem (P12); thus, we obtain $\overline{\text{C12}}\text{: }\left(1+\Gamma_{k}\right)\left|t_{k,k}\right|^{2} \geq \Gamma_{k}\left(\sum_{i\in\mathcal{K}}\left|t_{k,i}\right|^{2}+\hat{I}_{k}+\frac{\sigma_{k,c}^{2}}{\rho_{k}}\right)$. Then, we define $\text{C21: }z_{k}\geq 0$ as $z_{k}\triangleq\left(1+\Gamma_{k}\right)\left|t_{k,k}\right|^{2}-\Gamma_{k}\left(\sum_{i\in\mathcal{K}}\left|t_{k,i}\right|^{2}+\hat{I}_{k}\right)$, such that $\text{C22: }z_{k}+\Gamma_{k}\left(\sum_{i\in\mathcal{K}}\left|t_{k,i}\right|^{2}+\hat{I}_{k}\right)=\left(1+\Gamma_{k}\right)\left|t_{k,k}\right|^{2}$ and $\text{C23: }z_{k}\rho_{k}\geq \Gamma_{k}\sigma_{k,c}^{2}$. Hence, we cast (P12) into the following convex problem:
\begin{equation}\label{eq:BCD8}
\text{(P13):}\ \!\underset{\left\{\rho_{k}\right\},\left\{t_{k,i}\right\}}{\min}\ \sum_{k\in\mathcal{K}}\sum_{i\in\mathcal{K}}\left|\mathbf{h}_{k}^{\dagger}\mathbf{w}_{i}-t_{k,i}\right|^{2} \text{ s.t. } \text{C13, C19, C21--C23}.
\end{equation}
Problem (P13) is equivalent to an second order cone program (SOCP) formulation and it can be efficiently solved by software tools such as Gurobi~\cite{Gurobi}.

On the other hand, when all blocks except $\left\{\psi_{u,k}\right\}$ are fixed, problem (P9) is reduced (by ignoring the constant terms) to:
\begin{equation}\label{eq:BCD9}
\text{(P14):}\ \!\underset{\left\{\psi_{u,k}\right\}}{\min}\ \sum_{u\in\mathcal{U}}\sum_{k\in\mathcal{K}}\left|\mathbf{v}_{d,u}^{\dagger}\mathbf{w}_{k}-\psi_{u,k}\right|^{2} \text{ s.t. C14}. 
\end{equation} 
Similarly, when all blocks except $\left\{\lambda_{u,j}\right\}$ are fixed, problem (P9) is reduced (by ignoring the constant terms) to:
\begin{equation}\label{eq:BCD10}
\text{(P15):}\ \!\underset{\left\{\lambda_{u,j}\right\}}{\min}\ \sum_{u\in\mathcal{U}}\sum_{j\in\mathcal{U}}\left|\bm{\upsilon}^{\dagger}\mathbf{G}_{u}\mathbf{f}_{j}-\lambda_{u,j}\right|^{2} \text{ s.t. C15}. 
\end{equation} 
Problems (P14) and (P15) are quadratically constrained quadratic programs (QCQP) with a single constraint and they can be solved by applying the Lagrange duality and bisection methods, as in~\cite{IRSSWIPT2}. We omit the details here due to the limitation in pages.

\subsubsection{Update of the Penalty Coefficient}\label{subsubsec:3.3.5}
The TB and RB weights, PS ratios, and auxiliary variables are updated in the inner-layer of the proposed penalty-based BCD algorithm. In the outer-layer, the penalty coefficient $\omega$, which is initialized at a large value, is gradually decreased in each iteration according to $\omega^{(t+1)} = c\omega^{(t)}$, where $t$ denotes the outer-layer iteration index and $0 < c < 1$ is a scaling parameter, such that the transmit sum-power is minimized while we ensure that the equality constraints of problem (P8) are satisfied (within a predefined accuracy).

\subsubsection{Convergence and Computational Complexity}\label{subsubsec:3.3.6}
The overall algorithm is described in Alg.~\ref{algo:algoBCD}, where $\epsilon_{1}$ and $\epsilon_{2}$ are predefined convergence thresholds and 
\begin{equation*}
\xi\triangleq\max\left\{\left|\mathbf{h}_{k}^{\dagger}\mathbf{w}_{i}-t_{k,i}\right|^{2},\left|\mathbf{v}_{d,u}^{\dagger}\mathbf{w}_{k}-\psi_{u,k}\right|^{2},\left|\bm{\upsilon}^{\dagger}\mathbf{G}_{u}\mathbf{f}_{j}-\lambda_{u,j}\right|^{2}\right\}, \ \forall k,i\in\mathcal{K}, \ \forall u,j\in\mathcal{U},
\end{equation*}
denotes the stopping criterion. Note that Alg.~\ref{algo:algo1} is guaranteed to converge to a stationary point of (P1). The complexity of Alg.~\ref{algo:algoBCD} is $\mathcal{O}\left(I_{O}I_{I}\left(\mathcal{O}_{1}+\mathcal{O}_{2}+\mathcal{O}_{3}+\mathcal{O}_{4}\right)\right)$, where $I_{O}$ and $I_{I}$ represent the iteration time of the outer and inner loop, respectively, $\mathcal{O}_{1}=\left(K+U\right)\left(N^{2}+MN+M^{2}\right)+M^{2}$, $\mathcal{O}_{2}=I_{\bm{\upsilon}}K^{2}N$ where $I_{\bm{\upsilon}}$ corresponds to the iteration time of the RCG algorithm, $\mathcal{O}_{3}=\mathcal{O}(3K^{3.5})$, and $\mathcal{O}_{4}=\log\left(1/\epsilon_{3}\right)\left(U^{2}\left(K+1\right)+KU\right)$, where $\epsilon_{3}$ is the accuracy of bisection search.  
\begin{algorithm}[!t]
\centering \footnotesize
  \begin{algorithmic}[1]
  \State Initialize $\left\{\mathbf{w}_{k}\right\}$, $\bm{\upsilon}$, $\left\{\rho_{k}\right\}$, $\left\{t_{k,i}\right\}$, $\left\{\psi_{u,k}\right\}$, $\left\{\lambda_{u,j}\right\}$, and $\omega$; set $\epsilon_{1}>0$ and $\epsilon_{2}>0$. 
  \Repeat
  \Repeat
  	\State Update $\left\{\mathbf{w}_{k}\right\}$ by Eq.~(\ref{eq:BCD4}).
	\State Update $\bm{\upsilon}$ by solving (P11).  
	\State Update $\left\{\rho_{k}\right\}$ and $\left\{t_{k,i}\right\}$ by solving (P13).
	\State Update $\left\{\psi_{u,k}\right\}$ by solving (P14) and $\left\{\lambda_{u,j}\right\}$  by solving (P15).	
  \Until{\State The fractional decrease of the objective value in (P8) is below $\epsilon_{1}$.} 
  \State Update $\omega$ as $\omega_{\text{new}} = c\omega_{\text{old}}$.
  \Until {\State The stopping indicator $\xi$ is below $\epsilon_{2}$.}
\end{algorithmic}
\caption{BCD Algorithm for Solving Problem (P1).}
\label{algo:algoBCD}
\end{algorithm}

\subsection{Discrete IRS Phase Shifts}\label{subsec:3.4}
In practice, the reflection phase shift of each IRS element takes only discrete values from a finite set $\mathcal{F}$ with $F$ discrete levels. Let $\bar{\theta}_{n}$ be the discrete phase shift of the $n$-th IRS element and $\mathcal{F}\triangleq\left\{\tilde{\theta}_{1},\dots,\tilde{\theta}_{F}\right\}\triangleq\left\{0,2\pi/F,\dots,\left(F-1\right)2\pi/F\right\}$. In order to obtain a near-optimal discrete solution $\bar{\theta}_{n}^{\star}$ with low complexity, we map the (near-)optimal continuous solution $\theta_{n}$ to the nearest possible discrete level: $\bar{\theta}_{n}=\operatorname{arg} \ \underset{\tilde{\theta}_{f}\in\mathcal{F}}{\min} \ \left|\theta_{n}-\tilde{\theta}_{f}\right|$, $\forall n$, $\forall f=1,\dots,F$.      

\section{Robust Beamforming Design for Imperfect CSI}\label{sec:4}

\subsection{Channel Uncertainty and Statistical CSI Error Model}\label{subsec:4.1}
Under the occurrence of channel estimation errors, the actual direct and IRS-cascaded ST--SR $k$ and ST--PR $u$ channels as well as the actual IRS-cascaded PT--PR $u$ channels are given by:  
\begin{subequations}\label{eq:36}
\begin{alignat}{2}
&\mathbf{h}_{d,k} = \hat{\mathbf{h}}_{d,k} + \Delta\mathbf{h}_{d,k}; \ \mathbf{H}_{k} = \widehat{\mathbf{H}}_{k} + \Delta\mathbf{H}_{k}, \ \forall k, \label{eq:36a} \\
&\mathbf{v}_{d,u} = \hat{\mathbf{v}}_{d,u} + \Delta\mathbf{v}_{d,u}; \ \mathbf{V}_{u} = \widehat{\mathbf{V}}_{u} + \Delta\mathbf{V}_{u}, \ \forall u, \label{eq:36b} \\
&\mathbf{G}_{u} = \widehat{\mathbf{G}}_{u} + \Delta\mathbf{G}_{u}, \ \forall u, \label{eq:36c}
\end{alignat}
\end{subequations}
where $\left\{\hat{\mathbf{h}}_{d,k},\widehat{\mathbf{H}}_{k},\hat{\mathbf{v}}_{d,u},\widehat{\mathbf{V}}_{u},\widehat{\mathbf{G}}_{u}\right\}$ represent the corresponding channel estimates, which are known at the ST, and $\left\{\Delta\mathbf{h}_{d,k},\Delta\mathbf{H}_{k},\Delta\mathbf{v}_{d,u},\Delta\mathbf{V}_{u},\Delta\mathbf{G}_{u}\right\}$ denote the respective unknown channel estimation errors, which capture the channel uncertainty. Under a statistical CSI error model, we have $\Delta\mathbf{h}_{d,k}\sim\mathcal{CN}\left(\mathbf{0}_{M},\mathbf{C}_{\mathbf{h}_{d,k}}\right)$, $\operatorname{vec}\left(\Delta\mathbf{H}_{k}\right)\sim\mathcal{CN}\left(\mathbf{0}_{NM},\mathbf{C}_{\mathbf{H}_{k}}\right)$, $\Delta\mathbf{v}_{d,u}\sim\mathcal{CN}\left(\mathbf{0}_{M},\mathbf{C}_{\mathbf{v}_{d,u}}\right)$, $\operatorname{vec}\left(\Delta\mathbf{V}_{u}\right)\sim\mathcal{CN}\left(\mathbf{0}_{NM},\mathbf{C}_{\mathbf{V}_{u}}\right)$, and $\operatorname{vec}\left(\Delta\mathbf{G}_{u}\right)\sim\mathcal{CN}\left(\mathbf{0}_{NL},\mathbf{C}_{\mathbf{G}_{u}}\right)$, where $\mathbf{C}_{\mathbf{h}_{d,k}}\in\mathbb{H}^{M}$, $\mathbf{C}_{\mathbf{H}_{k}}\in\mathbb{H}^{NM}$, $\mathbf{C}_{\mathbf{v}_{d,u}}\in\mathbb{H}^{M}$, $\mathbf{C}_{\mathbf{V}_{u}}\in\mathbb{H}^{NM}$, and $\mathbf{C}_{\mathbf{G}_{u}}\in\mathbb{H}^{NL}$ denote the corresponding PSD error covariance matrices, which are known at the ST. 

\subsection{Problem Formulation}\label{subsec:4.2}
We consider the special case where we fix the receive PS ratios $\left\{\rho_{k}\right\}$ as $\rho_{k} = \sqrt{\omega_{R}\Gamma_{k}}/\left(\sqrt{\omega_{R}\Gamma_{k}}+\sqrt{\omega_{E}Q_{k}}\right)$, where $\omega_{R},\omega_{E}\in(0,1]$ denote arbitrary SINR and EH weights, respectively\footnote{The proposed approach applies to the general case where the receive PS ratios are not fixed as well, with the only required addition being an extra Taylor series approximation step.}~\cite{SLP}. In order to incorporate the channel uncertainty in our study, we introduce probabilistic QoS and interference constraints. Thus, we obtain:  
\begin{subequations}\label{eq:37}
\begin{alignat}{2}
&&& \text{(P16):}\ \!\underset{\left\{\mathbf{w}_{k}\right\},\bm{\upsilon}}{\min} \ \bar{\tau}\sum_{k\in\mathcal{K}}\left\|\mathbf{w}_{k}\right\|^{2} \label{eq:37a} \\
&&& \text{s.t.} \ \ \ \text{C24: }\operatorname{Pr}\left(\mathrm{SINR}_{k} \geq \Gamma_{k}\right) \geq 1-p_{k}, \ \forall k, \label{eq:37b} \\
&&& \ \ \ \ \ \ \ \text{C25: }\operatorname{Pr}\left(\bar{\tau}\left(1-\rho_{k}\right)E_{k}^{\text{In}} \geq Q_{k}\right) \geq 1-q_{k}, \ \forall k, \label{eq:37c} \\
&&& \ \ \ \ \ \ \ \text{C26: }\operatorname{Pr}\left(\mathrm{FISI}_{u} \leq E_{u}^{I}\right) \geq 1- \varsigma_{u}, \ \forall u, \label{eq:37d} \\
&&& \ \ \ \ \ \ \ \text{C27: }\operatorname{Pr}\left(\mathrm{CIUSI}_{u} \leq E_{\text{Th}}\right) \geq 1- \varrho_{u}, \ \forall u, \label{eq:37e} \\
&&& \ \ \ \ \ \ \ \text{C6: }\left|\bm{\upsilon}(n)\right|^{2} = 1, \ \forall n, \label{eq:37g}
\end{alignat}
\end{subequations}
where $p_{k},q_{k},\varsigma_{u},\varrho_{u}\in(0,1]$ are the corresponding maximum tolerable outage probabilities. In order to tackle this challenging optimization problem with non-convex probabilistic constraints, we follow the SDR-based AM approach presented in Sec.~\ref{sec:3} and adopt the BTI method. 

\subsection{TB Optimization}\label{subsec:4.3}
First, we turn our attention into TB optimization for fixed RB weights:
\begin{equation}\label{eq:38}
\text{(P17):}\ \!\underset{\left\{\mathbf{w}_{k}\right\}}{\min} \ \bar{\tau}\sum_{k\in\mathcal{K}}\left\|\mathbf{w}_{k}\right\|^{2} \  \text{s.t.} \ \text{C24--C26}.
\end{equation}  

By introducing the rank-one PSD matrix variables $\mathbf{W}_{k} \triangleq \mathbf{w}_{k}\mathbf{w}_{k}^{\dagger}\in\mathbb{H}^{M}$, we can replace the objective function in (P16) by the one in Eq.~(\ref{eq:12a}). Also, we can rewrite the SINR event $\mathrm{SINR}_{k} \geq \Gamma_{k}$ in C24 as $\mathbf{h}_{k}^{\dagger}\mathbf{B}_{k}\mathbf{h}_{k} - I_{k} \geq 0$, i.e., 
\begin{equation}\label{eq:39}
\underbrace{\left(\mathbf{h}_{d,k}^{\dagger}+\bm{\upsilon}^{\dagger}\mathbf{H}_{k}\right)\mathbf{B}_{k}\left(\mathbf{h}_{d,k}+\mathbf{H}_{k}^{\dagger}\bm{\upsilon}\right)}_{A_{k}} - I_{k} \geq 0, \ \forall k, 
\end{equation}
where $\mathbf{B}_{k} \triangleq \left(1+\frac{1}{\Gamma_{k}}\right)\mathbf{W}_{k} - \mathbf{D}_{k}$ with $\mathbf{D}_{k} \triangleq \sum_{i\in\mathcal{K}\setminus \{k\}}\mathbf{W}_{i}$. By using Eq.~(\ref{eq:36a}), $A_{k}$ in Eq.~(\ref{eq:39}) is expressed as:
\begin{align}\label{eq:40}
A_{k} = &\underbrace{\left(\hat{\mathbf{h}}_{d,k}^{\dagger}+\bm{\upsilon}^{\dagger}\widehat{\mathbf{H}}_{k}\right)\mathbf{B}_{k}\left(\hat{\mathbf{h}}_{d,k}+\widehat{\mathbf{H}}_{k}^{\dagger}\bm{\upsilon}\right)}_{B_{k}} + 2\text{Re}\left\{\underbrace{\left(\hat{\mathbf{h}}_{d,k}^{\dagger}+\bm{\upsilon}^{\dagger}\widehat{\mathbf{H}}_{k}\right)\mathbf{B}_{k}\left(\Delta\mathbf{h}_{d,k}+\Delta\mathbf{H}_{k}^{\dagger}\bm{\upsilon}\right)}_{C_{k}}\right\} \nonumber \\
&+ \underbrace{\left(\left(\Delta\mathbf{h}_{d,k}\right)^{\dagger}+\bm{\upsilon}^{\dagger}\Delta\mathbf{H}_{k}\right)\mathbf{B}_{k}\left(\Delta\mathbf{h}_{d,k}+\left(\Delta\mathbf{H}_{k}\right)^{\dagger}\bm{\upsilon}\right)}_{D_{k}}, \ \forall k.
\end{align}  
Let $\Delta\mathbf{h}_{d,k} = \mathbf{C}_{\mathbf{h}_{d,k}}^{1/2}\mathbf{i}_{\mathbf{h}_{d,k}}$ and $\operatorname{vec}\left(\Delta\mathbf{H}_{k}\right) = \mathbf{C}_{\mathbf{H}_{k}}^{1/2}\mathbf{i}_{\mathbf{H}_{k}}$, where $\mathbf{i}_{\mathbf{h}_{d,k}}\sim\mathcal{CN}\left(\mathbf{0}_{M},\mathbf{I}_{M}\right)$ and $\mathbf{i}_{\mathbf{H}_{k}}\sim\mathcal{CN}\left(\mathbf{0}_{NM},\mathbf{I}_{NM}\right)$. Let also $\widetilde{N} \triangleq M\widebar{N} = M(N+1)$. Then, by exploiting the property $\mathbf{X}^{\dagger} = \mathbf{X}$ of any covariance matrix $\mathbf{X}$, we can rewrite the term $C_{k}$ in Eq.~(\ref{eq:40}) as
\begin{equation}\label{eq:41}
C_{k} = \left(\hat{\mathbf{h}}_{d,k}^{\dagger}+\bm{\upsilon}^{\dagger}\widehat{\mathbf{H}}_{k}\right)\mathbf{B}_{k}\mathbf{C}_{\mathbf{h}_{d,k}}^{1/2}\mathbf{i}_{\mathbf{h}_{d,k}} + \operatorname{vec}^{T}\left(\bm{\upsilon}\left(\hat{\mathbf{h}}_{d,k}^{\dagger}+\bm{\upsilon}^{\dagger}\widehat{\mathbf{H}}_{k}\right)\mathbf{B}_{k}\right)\left(\mathbf{C}_{\mathbf{H}_{k}}^{1/2}\right)^{*}\mathbf{i}_{\mathbf{H}_{k}}^{*} = \mathbf{e}_{k}^{\dagger}\mathbf{i}_{k}, \ \forall k,
\end{equation}
where $\mathbf{i}_{k}\in\mathbb{C}^{\widetilde{N}}$ is defined as $\mathbf{i}_{k} \triangleq \left[\mathbf{i}_{\mathbf{h}_{d,k}}^{\dagger} \ \mathbf{i}_{\mathbf{H}_{k}}^{T}\right]^{\dagger}$ and $\mathbf{e}_{k}\in\mathbb{C}^{\widetilde{N}}$ is given by
\begin{equation}\label{eq:42}
\mathbf{e}_{k} \triangleq \begin{bmatrix} \mathbf{C}_{\mathbf{h}_{d,k}}^{1/2}\mathbf{B}_{k}\left(\hat{\mathbf{h}}_{d,k}+\widehat{\mathbf{H}}_{k}^{\dagger}\bm{\upsilon}\right)\\ \left(\mathbf{C}_{\mathbf{H}_{k}}^{1/2}\right)^{T}\operatorname{vec}^{*}\left(\bm{\upsilon}\left(\hat{\mathbf{h}}_{d,k}^{\dagger}+\bm{\upsilon}^{\dagger}\widehat{\mathbf{H}}_{k}\right)\mathbf{B}_{k}\right)\end{bmatrix}, \ \forall k.
\end{equation}
Similarly, by defining the PSD matrix $\bm{\Xi} \triangleq \bm{\upsilon}\bm{\upsilon}^{\dagger}\in\mathbb{H}^{N}$, we have
\begin{align}\label{eq:43}
D_{k} = &\mathbf{i}_{\mathbf{h}_{d,k}}^{\dagger}\mathbf{C}_{\mathbf{h}_{d,k}}^{1/2}\mathbf{B}_{k}\mathbf{C}_{\mathbf{h}_{d,k}}^{1/2}\mathbf{i}_{\mathbf{h}_{d,k}} + 2\text{Re}\left\{\mathbf{i}_{\mathbf{h}_{d,k}}^{\dagger}\mathbf{C}_{\mathbf{h}_{d,k}}^{1/2}\left(\mathbf{B}_{k}\otimes\bm{\upsilon}^{T}\right)\left(\mathbf{C}_{\mathbf{H}_{k}}^{1/2}\right)^{*}\mathbf{i}_{\mathbf{H}_{k}}^{*}\right\} \nonumber \\
&+ \mathbf{i}_{\mathbf{H}_{k}}^{T}\mathbf{C}_{\mathbf{H}_{k}}^{1/2}\left(\mathbf{B}_{k}\otimes\bm{\Xi}^{T}\right)\left(\mathbf{C}_{\mathbf{H}_{k}}^{1/2}\right)^{*}\mathbf{i}_{\mathbf{H}_{k}}^{*} = \mathbf{i}_{k}^{\dagger}\mathbf{E}_{k}\mathbf{i}_{k}, \ \forall k,
\end{align}
where $\mathbf{E}_{k}\in\mathbb{C}^{\widetilde{N}\times \widetilde{N}}: \mathbf{E}_{k}\triangleq\left[\mathbf{M}_{1},\mathbf{M}_{2};\mathbf{M}_{3},\mathbf{M}_{4}\right]$, with $\mathbf{M}_{1}\in\mathbb{C}^{M}$, $\mathbf{M}_{2}\in\mathbb{C}^{M\times NM}$, $\mathbf{M}_{3}\in\mathbb{C}^{NM\times M}$, and $\mathbf{M}_{4}\in\mathbb{C}^{NM\times NM}$ defined as   
\begin{subequations}\label{eq:44}
\begin{alignat}{2}
&\mathbf{M}_{1} \triangleq \mathbf{C}_{\mathbf{h}_{d,k}}^{1/2}\mathbf{B}_{k}\mathbf{C}_{\mathbf{h}_{d,k}}^{1/2}; \ \mathbf{M}_{2} \triangleq \mathbf{C}_{\mathbf{h}_{d,k}}^{1/2}\left(\mathbf{B}_{k}\otimes\bm{\upsilon}^{T}\right)\left(\mathbf{C}_{\mathbf{H}_{k}}^{1/2}\right)^{*}, \ \forall k, \label{eq:44b} \\
&\mathbf{M}_{3} \triangleq \left(\mathbf{C}_{\mathbf{H}_{k}}^{1/2}\right)^{T}\left(\mathbf{B}_{k}\otimes\bm{\upsilon}^{*}\right)\mathbf{C}_{\mathbf{h}_{d,k}}^{1/2}; \ \mathbf{M}_{4} \triangleq \mathbf{C}_{\mathbf{H}_{k}}^{1/2}\left(\mathbf{B}_{k}\otimes\bm{\Xi}^{T}\right)\left(\mathbf{C}_{\mathbf{H}_{k}}^{1/2}\right)^{*}, \ \forall k. \label{eq:44d}
\end{alignat}
\end{subequations}
Finally, by denoting $e_{k}\triangleq B_{k} - I_{k}$, the probabilistic SINR constraints C24 become
\begin{equation}\label{eq:45}
\text{C28: }\operatorname{Pr}\left(\mathbf{i}_{k}^{\dagger}\mathbf{E}_{k}\mathbf{i}_{k} + 2\text{Re}\left\{\mathbf{e}_{k}^{\dagger}\mathbf{i}_{k}\right\} + e_{k} \geq 0\right) \geq 1-p_{k}, \ \forall k.
\end{equation}

These semi-definite probabilistic constraints are still intractable. In order to obtain a convex approximation of them, we rely on the BTI approach~\cite{BTI}. Specifically, by introducing the auxiliary variables $\dot{\mathbf{x}} = \left[\dot{x}_{1},\dots,\dot{x}_{K}\right]^{T}$ and $\dot{\mathbf{y}} = \left[\dot{y}_{1},\dots,\dot{y}_{K}\right]^{T}$, we obtain:
\begin{subequations}\label{eq:49}
\begin{alignat}{2}
&\text{C28a: }\operatorname{Tr}\left(\mathbf{E}_{k}\right)-\sqrt{2\text{ln}\left(1/p_{k}\right)}\dot{x}_{k} + \text{ln}\left(p_{k}\right)\dot{y}_{k} + e_{k} \geq 0, \ \forall k, \label{eq:49a} \\
&\text{C28b: } \left\|\begin{bmatrix}
\operatorname{vec}\left(\mathbf{E}_{k}\right) \\ \sqrt{2}\mathbf{e}_{k}
\end{bmatrix}\right\| \leq \dot{x}_{k}, \ \forall k, \label{eq:49b} \\
&\text{C28c: } \dot{y}_{k}\mathbf{I}_{\widetilde{N}} + \mathbf{E}_{k} \succeq \mathbf{0}, \ \text{C28d: }\dot{y}_{k} \geq 0, \ \forall k. \label{eq:49c}
\end{alignat}
\end{subequations}
Let's assume, for the convenience of derivations, that $\mathbf{C}_{\mathbf{h}_{d,k}}=\varepsilon_{\mathbf{h}_{d,k}}^{2}\mathbf{I}_{M}$ and $\mathbf{C}_{\mathbf{H}_{k}}=\varepsilon_{\mathbf{H}_{k}}^{2}\mathbf{I}_{NM}$. Then, after some mathematical manipulations, the constraints C28a--C28c are simplified as:
\begin{subequations}\label{eq:Add8}
\begin{alignat}{2}
&\overline{\text{C28a}}\text{: }\left(\varepsilon_{\mathbf{h}_{d,k}}^{2}+\varepsilon_{\mathbf{H}_{k}}^{2}N\right)\operatorname{Tr}\left(\mathbf{B}_{k}\right) -\sqrt{2\text{ln}\left(1/p_{k}\right)}\dot{x}_{k} + \text{ln}\left(p_{k}\right)\dot{y}_{k} + e_{k} \geq 0, \ \forall k, \label{eq:Add8a} \\
&\overline{\text{C28b}}\text{: } \left\| \begin{bmatrix}
\left(\varepsilon_{\mathbf{h}_{d,k}}^{2}+\varepsilon_{\mathbf{H}_{k}}^{2}N\right)\operatorname{vec}\left(\mathbf{B}_{k}\right) \\ \sqrt{2\left(\varepsilon_{\mathbf{h}_{d,k}}^{2}+\varepsilon_{\mathbf{H}_{k}}^{2}N\right)}\mathbf{B}_{k}\left(\hat{\mathbf{h}}_{d,k}+\widehat{\mathbf{H}}_{k}^{\dagger}\bm{\upsilon}\right)
\end{bmatrix}\right\| \leq \dot{x}_{k}, \ \forall k, \label{eq:Add8b} \\
&\overline{\text{C28c}}\text{: }\dot{y}_{k}\mathbf{I}_{M} + \left(\varepsilon_{\mathbf{h}_{d,k}}^{2}+\varepsilon_{\mathbf{H}_{k}}^{2}N\right)\mathbf{B}_{k} \succeq \mathbf{0}, \ \forall k. \label{eq:Add11a}
\end{alignat}
\end{subequations} 

Similar to the above, we can recast the EH event in C25 as $\mathbf{h}_{k}^{\dagger}\bm{\Sigma}_{k}\mathbf{h}_{k} + \tilde{I}_{k} \geq 0$, where $\bm{\Sigma}_{k} \triangleq \left(\bar{\tau}\left(1-\rho_{k}\right)/Q_{k}\right)$ $\sum_{i\in\mathcal{K}}\mathbf{W}_{i}$. Thus, we approximate the probabilistic EH constraints C25 as:
\begin{subequations}\label{eq:Add9}
\begin{alignat}{2}
&\overline{\text{C29a}}\text{: }\left(\varepsilon_{\mathbf{h}_{d,k}}^{2}+\varepsilon_{\mathbf{H}_{k}}^{2}N\right)\operatorname{Tr}\left(\bm{\Sigma}_{k}\right)-\sqrt{2\text{ln}\left(1/q_{k}\right)}\tilde{x}_{k} + \text{ln}\left(q_{k}\right)\tilde{y}_{k} + \tilde{e}_{k} \geq 0, \ \forall k, \label{eq:Add9a} \\
&\overline{\text{C29b}}\text{: }\left\| \begin{bmatrix}
\left(\varepsilon_{\mathbf{h}_{d,k}}^{2}+\varepsilon_{\mathbf{H}_{k}}^{2}N\right)\operatorname{vec}\left(\bm{\Sigma}_{k}\right) \\ \sqrt{2\left(\varepsilon_{\mathbf{h}_{d,k}}^{2}+\varepsilon_{\mathbf{H}_{k}}^{2}N\right)}\bm{\Sigma}_{k}\left(\hat{\mathbf{h}}_{d,k}+\widehat{\mathbf{H}}_{k}^{\dagger}\bm{\upsilon}\right)
\end{bmatrix}\right\| \leq \tilde{x}_{k}, \ \forall k, \label{eq:Add9b} \\
&\overline{\text{C29c}}\text{: } \tilde{y}_{k}\mathbf{I}_{M} + \left(\varepsilon_{\mathbf{h}_{d,k}}^{2}+\varepsilon_{\mathbf{H}_{k}}^{2}N\right)\bm{\Sigma}_{k} \succeq \mathbf{0}, \ \overline{\text{C29d}}\text{: } \tilde{y}_{k} \geq 0, \forall k, \label{eq:Add11b}
\end{alignat}
\end{subequations}
where $\tilde{e}_{k} = \left(\hat{\mathbf{h}}_{k,d}^{\dagger}+\bm{\upsilon}^{\dagger}\widehat{\mathbf{H}}_{k}\right)\bm{\Sigma}_{k}\left(\hat{\mathbf{h}}_{k,d}+\widehat{\mathbf{H}}_{k}^{\dagger}\bm{\upsilon}\right) + \hat{I}_{k}$ while $\tilde{\mathbf{x}} = \left[\tilde{x}_{1},\dots,\tilde{x}_{K}\right]^{T}$ and $\tilde{\mathbf{y}} = \left[\tilde{y}_{1},\dots,\tilde{y}_{K}\right]^{T}$ are auxiliary variables.

Furthermore, the FISI event in C26 can be rewritten as $\mathbf{v}_{u}^{\dagger}\bm{\Lambda}_{k}\mathbf{v}_{u} - E_{u}^{(I)} \leq 0$, where $\bm{\Lambda}_{k} \triangleq \bar{\tau}\sum_{k\in\mathcal{K}}\mathbf{W}_{k}$. Therefore, by assuming for convenience and without loss of generality that $\mathbf{C}_{\mathbf{v}_{d,u}}=\varepsilon_{\mathbf{v}_{d,u}}^{2}\mathbf{I}_{M}$ and $\mathbf{C}_{\mathbf{V}_{u}}=\varepsilon_{\mathbf{V}_{u}}^{2}\mathbf{I}_{N}$, we obtain the following approximation for the probabilistic FISI constraints C26:
\begin{subequations}\label{eq:Add10}
\begin{alignat}{2}
&\overline{\text{C30a}}\text{: }\left(\varepsilon_{\mathbf{v}_{d,u}}^{2}+\varepsilon_{\mathbf{V}_{u}}^{2}N\right)\operatorname{Tr}\left(\bm{\Lambda}_{u}\right)-\sqrt{2\text{ln}\left(1/\varsigma_{u}\right)}\bar{x}_{u} + \text{ln}\left(\varsigma_{u}\right)\bar{y}_{u} + \bar{e}_{u} \leq 0, \ \forall u, \label{eq:Add10a} \\
&\overline{\text{C30b}}\text{: } \left\| \begin{bmatrix}
\left(\varepsilon_{\mathbf{v}_{d,u}}^{2}+\varepsilon_{\mathbf{V}_{u}}^{2}N\right)\operatorname{vec}\left(\bm{\Lambda}_{u}\right) \\ \sqrt{2\left(\varepsilon_{\mathbf{v}_{d,u}}^{2}+\varepsilon_{\mathbf{V}_{u}}^{2}N\right)}\bm{\Lambda}_{u}\left(\hat{\mathbf{v}}_{d,u}+\hat{\mathbf{V}}_{u}^{\dagger}\bm{\upsilon}\right)
\end{bmatrix}\right\| \leq \bar{x}_{u}, \ \forall u, \label{eq:Add10b} \\
&\overline{\text{C30c}}\text{: } \bar{y}_{u}\mathbf{I}_{M} + \left(\varepsilon_{\mathbf{v}_{d,u}}^{2}+\varepsilon_{\mathbf{V}_{u}}^{2}N\right)\bm{\Lambda}_{u} \succeq \mathbf{0}, \ \overline{\text{C30d}}\text{: } \bar{y}_{u} \geq 0, \ \forall u,\label{eq:Add11c}
\end{alignat}
\end{subequations}
where $\bar{e}_{u} \triangleq \left(\hat{\mathbf{v}}_{d,u}^{\dagger}+\bm{\upsilon}^{\dagger}\widehat{\mathbf{V}}_{u}\right)\bm{\Lambda}_{k}\left(\hat{\mathbf{v}}_{d,u}+\widehat{\mathbf{V}}_{u}^{\dagger}\bm{\upsilon}\right) - E_{u}^{I}$ whereas $\bar{\mathbf{x}} = \left[\bar{x}_{1},\dots,\bar{x}_{U}\right]^{T}$ and $\bar{\mathbf{y}} = \left[\bar{y}_{1},\dots,\bar{y}_{U}\right]^{T}$ represent auxiliary variables.

Finally, instead of dropping the constraints $\operatorname{Rank}\left(\mathbf{W}_{k}\right) = 1$, we follow the penalty-based SCA approach described in problem (P7). Thus, we obtain the following robust formulation of problem (P16):
\begin{subequations}\label{eq:Alt}
\begin{alignat}{2}
&&&\text{(P18):}\ \!\underset{\substack{\left\{\mathbf{W}_{k}\right\},\left\{\dot{x}_{k},\dot{y}_{k}\right\} \\ \left\{\tilde{x}_{k},\tilde{y}_{k}\right\},\left\{\bar{x}_{u},\bar{y}_{u}\right\}}}{\min} \ \sum_{k\in\mathcal{K}}\operatorname{Tr}\left(\mathbf{W}_{k}\right) + \delta\left(\left\|\mathbf{W}_{k}\right\|_{*}-\bar{W}_{k}^{(t)}\right) \label{eq:Alta} \\
&&&\text{s.t.} \ \ \ \overline{\text{C28a}}\text{--}\overline{\text{C28c}}, \ \text{C28d}, \ \overline{\text{C29a}}\text{--}\overline{\text{C29d}}, \ \overline{\text{C30a}}\text{--}\overline{\text{C30d}}, \label{eq:Altb}
\end{alignat}
\end{subequations}
where $\delta>0$ is a constant that penalizes the objective function for any matrix $\mathbf{W}_{k}$ with rank higher than one. Then, we can obtain $\mathbf{w}_{k}^{\star}$ via Cholesky decomposition. Notice that the solution of the approximated problem (P18) is a feasible, yet suboptimal solution of the original probabilistic problem (P17)~\cite{RobustOpt}.  

\subsection{RB Optimization}\label{subsec:4.4}
We proceed with the feasibility check problem associated with the reflection phase shifts, for given transmit precoding vectors. This problem is formulated accordingly:
\begin{equation}\label{eq:55}
\text{(P19):}\ \text{Find} \ \bm{\upsilon} \ \text{s.t.} \ \text{C24--C27, C6}.
\end{equation}  

In order to solve problem (P19), we rely on the SDR- and BTI-based derivations obtained earlier. We also introduce the slack variables $\tau_{k}$, $\lambda_{k}$, $\mu_{u}$, and $\delta_{u}$ that are associated with the SINR, EH, FISI, and CIUSI constraints, respectively, to improve the convergence of the algorithm. We note that $e_{k}$, $\tilde{e}_{k}$, and $\bar{e}_{u}$ are non-concave in $\bm{\upsilon}$. Thus, we utilize the first-order Taylor inequality given in~\cite{IRSCascCSI1} to obtain linear approximations of these expressions:
\begin{subequations}\label{eq:Add7}
\begin{alignat}{2}
e_{k} \approx & 2\operatorname{Re}\left\{\bm{\upsilon}^{\dagger}\widehat{\mathbf{H}}_{k}\mathbf{B}_{k}\widehat{\mathbf{H}}_{k}^{\dagger}\bm{\upsilon}\right\} - \bm{\upsilon}^{\dagger}\widehat{\mathbf{H}}_{k}\mathbf{B}_{k}\widehat{\mathbf{H}}_{k}^{\dagger}\bm{\upsilon} - \bm{\upsilon}^{\dagger}\widehat{\mathbf{H}}_{k}\mathbf{D}_{k}\mathbf{D}_{k}^{\dagger}\widehat{\mathbf{H}}_{k}^{\dagger}\bm{\upsilon} \nonumber \\
&+ 2\operatorname{Re}\left\{\bm{\upsilon}^{\dagger}\widehat{\mathbf{H}}_{k}\mathbf{B}_{k}\hat{\mathbf{h}}_{k}\right\} + \hat{\mathbf{h}}_{k}^{\dagger}\mathbf{B}_{k}\hat{\mathbf{h}}_{k} - I_{k} - \tau_{k}, \ \forall k, \label{eq:Add7a} \\
\tilde{e}_{k}\approx & 2\operatorname{Re}\left\{\bm{\upsilon}^{\dagger}\widehat{\mathbf{H}}_{k}\bm{\Sigma}_{i}\widehat{\mathbf{H}}_{k}^{\dagger}\bm{\upsilon}\right\} - \bm{\upsilon}^{\dagger}\widehat{\mathbf{H}}_{k}\bm{\Sigma}_{i}\widehat{\mathbf{H}}_{k}^{\dagger}\bm{\upsilon} + 2\operatorname{Re}\left\{\bm{\upsilon}^{\dagger}\widehat{\mathbf{H}}_{k}\bm{\Sigma}_{i}\hat{\mathbf{h}}_{k}\right\} + \hat{\mathbf{h}}_{k}^{\dagger}\bm{\Sigma}_{i}\hat{\mathbf{h}}_{k} + \hat{I}_{k} - \lambda_{k}, \ \forall k, \label{eq:Add7b} \\
\bar{e}_{u} \approx & 2\operatorname{Re}\left\{\bm{\upsilon}^{\dagger}\widehat{\mathbf{V}}_{u}\bm{\Lambda}_{k}\widehat{\mathbf{V}}_{u}^{\dagger}\bm{\upsilon}\right\} - \bm{\upsilon}^{\dagger}\widehat{\mathbf{V}}_{u}\bm{\Lambda}_{k}\widehat{\mathbf{V}}_{u}^{\dagger}\bm{\upsilon} + 2\operatorname{Re}\left\{\bm{\upsilon}^{\dagger}\widehat{\mathbf{V}}_{u}\bm{\Lambda}_{k}\hat{\mathbf{v}}_{u}\right\} + \hat{\mathbf{v}}_{u}^{\dagger}\bm{\Lambda}_{k}\hat{\mathbf{v}}_{u} - E_{u}^{I} + \mu_{u}, \ \forall u. \label{eq:Add7c}
\end{alignat}
\end{subequations} 
We also notice that the constraints $\overline{\text{C28c}}$--$\overline{\text{C30c}}$ are independent of $\bm{\upsilon}$. 

Next, we have to obtain a safe approximation of the probabilistic CIUSI constraints C27. The CIUSI event is written as
\begin{equation}\label{eq:57}
\underbrace{\sum_{j\in\mathcal{U}}\bm{\upsilon}^{\dagger}\mathbf{G}_{u}\mathbf{f}_{j}\mathbf{f}_{j}^{\dagger}\mathbf{G}_{u}^{\dagger}\bm{\upsilon}}_{E_{u}} - E_{\text{Th}} \leq 0, \ \forall u.
\end{equation}
Let us define $\mathbf{F}_{j}\in\mathbb{H}^{L}$ and $\widebar{\mathbf{F}}\in\mathbb{C}^{L\times L}$ as $\mathbf{F}_{j}\triangleq\mathbf{f}_{j}\mathbf{f}_{j}^{\dagger}$ and $\widebar{\mathbf{F}}\triangleq\sum_{j\in\mathcal{U}}\mathbf{F}_{j}$, respectively. Then, we can express $E_{u}$ in Eq.~(\ref{eq:57}) as
\begin{equation}\label{eq:58}
E_{u} = \underbrace{\bm{\upsilon}^{\dagger}\widehat{\mathbf{G}}_{u}\widebar{\mathbf{F}}\widehat{\mathbf{G}}_{u}^{\dagger}\bm{\upsilon}}_{F_{u}} + 2\operatorname{Re}\left\{\underbrace{\bm{\upsilon}^{\dagger}\widehat{\mathbf{G}}_{u}\widebar{\mathbf{F}}\left(\Delta\mathbf{G}_{u}\right)^{\dagger}\bm{\upsilon}}_{G_{u}}\right\} +\underbrace{\bm{\upsilon}^{\dagger}\Delta\mathbf{G}_{u}\widebar{\mathbf{F}}\left(\Delta\mathbf{G}_{u}\right)^{\dagger}\bm{\upsilon}}_{H_{u}}, \ \forall u.
\end{equation}
By using the BTI approach as in Sec.~\ref{subsec:4.3}, introducing the auxiliary variables $\hat{\mathbf{x}} = \left[\hat{x}_{1},\dots,\hat{x}_{U}\right]^{T}$ and $\hat{\mathbf{y}} = \left[\hat{y}_{1},\dots,\hat{y}_{U}\right]^{T}$, and assuming for the convenience of derivations that $\mathbf{C}_{\mathbf{G}_{u}}=\varepsilon_{\mathbf{G}_{u}}^{2}\mathbf{I}_{\widehat{N}}$, we safely approximate C27 as:
\begin{subequations}\label{eq:63}
\begin{alignat}{2}
&\overline{\text{C31a}}\text{: }\varepsilon_{\mathbf{G}_{u}}^{2}\operatorname{Tr}\left(\widebar{\mathbf{F}}\right)-\sqrt{2\text{ln}\left(1/\varrho_{u}\right)}\hat{x}_{u} + \text{ln}\left(\varrho_{u}\right)\hat{y}_{k} + \hat{e}_{u} \leq 0, \ \forall k, \label{eq:63a} \\
&\overline{\text{C31b}}\text{: }\left\| \begin{bmatrix}
\varepsilon_{\mathbf{G}_{u}}^{2}\operatorname{vec}\left(\widebar{\mathbf{F}}\right) \\ \sqrt{2N}\varepsilon_{\mathbf{G}_{u}}\widebar{\mathbf{F}}\left(\hat{\mathbf{G}}_{u}^{\dagger}\bm{\upsilon}\right)
\end{bmatrix}\right\| \leq \hat{x}_{u}, \ \forall u, \label{eq:63b} \\
&\overline{\text{C31c}}\text{: }\hat{y}_{u}\mathbf{I}_{L} + \varepsilon_{\mathbf{G}_{u}}^{2}\widebar{\mathbf{F}} \succeq \mathbf{0}, \ \overline{\text{C31d}}\text{: }\hat{y}_{u} \geq 0, \ \forall u,\label{eq:63c}
\end{alignat}
\end{subequations}
where 
\begin{equation}\label{eq:60}
\hat{e}_{u} \approx 2\operatorname{Re}\left\{\bm{\upsilon}^{\dagger}\widehat{\mathbf{G}}_{}\widebar{\mathbf{F}}\widehat{\mathbf{G}}_{u}^{\dagger}\bm{\upsilon}\right\} - \bm{\upsilon}^{\dagger}\widehat{\mathbf{G}}_{u}\widebar{\mathbf{F}}\widehat{\mathbf{G}}_{u}^{\dagger}\bm{\upsilon}, \ \forall u.
\end{equation}
and we note that $\overline{\text{C31c}}$ is independent of $\bm{\upsilon}$.

Finally, we utilize the penalty CCP method to handle the unit-modulus constraints in C6. Specifically, by introducing the slack variable $\bm{\zeta} = \left[\zeta_{1},\dots,\zeta_{2N}\right]^{T}$, we can equivalently write these constraints as follows~\cite{IRSCascCSI1}:
\begin{equation}\label{eq:Add12}
\text{C32a: }\left|\upsilon_{n}^{[r]}\right|^{2}-2\operatorname{Re}\left(\upsilon_{n}^{*}\upsilon_{n}^{[r]}\right) \leq \zeta_{n} - 1; \ \text{C32b: }\left|\upsilon_{n}\right|^{2} \leq 1 + \zeta_{N+n}; \ \text{C32c: } \bm{\zeta} \geq 0, \forall n,
\end{equation}
where Eq.~(\ref{eq:Add12}) is computed at fixed $\upsilon_{n}^{[r]}$, with $r$ denoting the iteration index for the penalty CCP algorithm. Hence, we can transform the feasibility check problem (P19) into:
\begin{subequations}\label{eq:xx}
\begin{alignat}{2}
&&& \text{(P20):}\ \!\underset{\substack{\bm{\upsilon},\left\{\tau_{k}\right\},\left\{\lambda_{k}\right\}, \\ \left\{\mu_{u}\right\},\left\{\delta_{u}\right\},\bm{\zeta}}}{\max} \ \sum_{k\in\mathcal{K}}\left(\tau_{k}+\lambda_{k}\right) - \sum_{u\in\mathcal{U}}\left(\mu_{u}+\delta_{u}\right) - \varpi^{[r]}\sum_{n\in\mathcal{N}}\zeta_{n} \label{eq:xxa} \\
&&& \text{s.t.} \ \ \ \overline{\text{C28a}}\text{--}\overline{\text{C31a}}, \ \overline{\text{C28b}}\text{--}\overline{\text{C31b}}, \ \text{C28d}, \ \overline{\text{C29d}}\text{--}\overline{\text{C31d}}, \ \overline{\text{C32a}}\text{--}\overline{\text{C32c}},   \label{eq:xxb}
\end{alignat}
\end{subequations}
where $\left\|\bm{\zeta}\right\|_{1}\triangleq\sum_{n\in\mathcal{N}}\zeta_{n}$ is the penalty term introduced in the objective function to ensure that the unit-modulus constraints will be satisfied and $\varpi^{[r]}$ is the corresponding regularization factor that allows us to avoid numerical problems. (P20) is a convex problem and it can be solved by CVX. The penalty CCP algorithm and the AM algorithm for solving problem (P16) are presented in Alg.~\ref{algo:algo2} and~\ref{algo:algo3}, respectively. Note that the penalty CCP algorithm restarts from a new initial point $\bm{\upsilon}^{[0]}$ if a feasible solution has not been found after $R_{\max}$ iterations.

The complexity of the considered problems is dominated by the respective SOC or/and linear matrix inequality (LMI) constraints. Hence, we can approximate it using the approach described in~\cite{IRSCascCSI1}. Specifically, the complexity of problems (P18) and (P20) for obtaining an $\epsilon$-accurate solution is approximated as 
\begin{equation*}
\mathcal{O}\left(\operatorname{ln}\left(1/\epsilon\right)\sqrt{a_{1}M+2b_{1}}n_{1}\left(n_{1}^{2}+n_{1}a_{1}M^{2}+a_{1}M^{3}+n_{1}b_{1}c_{1}^{2}\right)\right)
\end{equation*}
and 
\begin{equation*}
\mathcal{O}\left(\operatorname{ln}\left(1/\epsilon\right)\sqrt{2\left(N+b_{1}+U\right)}N\left(N^{2}+12N^{3}+Nb_{1}c_{1}^{2}+NUc_{2}^{2}\right)\right),
\end{equation*}
respectively, where $a_{1}\triangleq 3K+U$, $b_{1}\triangleq 2K+U$, $c_{1}\triangleq M\left(M+1\right)$, $c_{2}\triangleq L\left(L+1\right)$, and $n_{1}\triangleq KM$. We note that the robust beamforming design presents higher complexity than the schemes derived for perfect CSI. We should mention that in practice the iteration complexity is commonly much smaller than its worst-case value given here.
\begin{algorithm}[!t]
\centering \footnotesize
  \begin{algorithmic}[1]
  \State Set initial points $\bm{\upsilon}^{[0]}$, $\eta^{[0]}>1$, and iteration index $r = 0$.
  \Repeat
  \If {$r<R_{\max}$}
  	\State Update $\bm{\upsilon}^{[r+1]}$ from problem (P20).
  	\State $\varpi^{[r+1]} = \min\left\{\eta\varpi^{[r]},\varpi_{\max}\right\}$.
  	\State $r = r + 1$. 
  \Else
  	\State Set initial points $\bm{\upsilon}^{[0]}$ and $\eta^{[0]}>1$ and iteration index $r=0$.
  \EndIf    
  \Until{$\left|\bm{\zeta}\right|_{1}\leq \chi$ and $\left\|\bm{\upsilon}^{[r]}-\bm{\upsilon}^{[r-1]}\right\|_{1}\leq \nu$.}
  \State Output: $\bm{\upsilon}^{(t)}=\bm{\upsilon}^{[r]}$.
\end{algorithmic}
\caption{Penalty CCP Algorithm for Solving Problem (P20).}
\label{algo:algo2}
\end{algorithm}
\begin{algorithm}[!t]
\centering \footnotesize
  \begin{algorithmic}[1]
  \State Set initial point $\bm{\upsilon}^{(0)}$, tolerance threshold $0 \leq \epsilon \ll 1$, maximum number of iterations $T_{\max}$, and iteration index $t=1$.
  \Repeat
  	\State Solve (P18) with given $\bm{\upsilon}^{(t-1)}$ to obtain $\mathbf{w}_{k}^{(t)}$.
  	\State Solve (P20) with given $\mathbf{w}_{k}^{(t)}$ via Alg.~\ref{algo:algo2} to obtain $\bm{\upsilon}^{(t)}$.
  	\State Set: $t = t + 1$.    
  \Until{\State $\left|f_{t}-f_{t-1}\right|/\left|f_{t-1}\right| \leq \varepsilon $, where $f_{t} = \bar{\tau}\sum\limits_{k\in\mathcal{K}}\left\|\mathbf{w}_{k}^{(t)}\right\|^{2}$.}
  \State Output: $\left\{\mathbf{w}_{k}^{\star},\bm{\upsilon}^{\star}\right\} = \left\{\mathbf{w}_{k}^{(t)},\bm{\upsilon}^{(t)}\right\}$. 
\end{algorithmic}
\caption{AM Algorithm for Solving Problem (P16).}
\label{algo:algo3}
\end{algorithm}  

\section{Numerical Simulation Results}\label{sec:5}
In this section, we provide numerical simulation results to evaluate the proposed algorithms and study the impact of the various parameters on system performance. In the simulation setup, the ST, the PT, and the IRS are located at (0 m, 0 m), (0 m, -20 m), and (5 m, 5 m), respectively, whereas the SRs and the PRs are uniformly placed on circles with centers located at (5 m, 0 m) and (5 m, -20 m), respectively, and radius of 2 m each. In order to facilitate the interpretation of the results by the readers, we plot the transmit sum-power of the ST, which is obtained by dividing the achieved value of $E_{s}$ by $\bar{\tau}$, vs. various varying parameters. Under this spirit, we convert the harvested and interference energy as well as the corresponding thresholds to respective power thresholds in the same way. In this context, we assume that $M=L=8$, $N=64$, $K=U=2$, $\Gamma_{k} = \Gamma = 0$ dB and $Q_{k} = Q = -20$ dBm as well as $\sigma_{k}^{2}=\sigma^{2} = -70$ dBm and $\sigma_{k,c}^{2}=\sigma_{c}^{2} = -50$ dBm  $\forall k$, $E_{u}^{I} = E_{I} = 0$ dBm, $\forall u$, and $E_{\text{Th}}=10^{-6}$, unless it is explicitly mentioned otherwise. We set the path loss exponents of the transmitter--IRS and IRS--user links to $\alpha_{\text{IRS}} = 2.2$ and that of the transmitter--receiver links to $\alpha_{\text{RX}}=3.6$, while the mean path loss is set to $C_{0} = 30$ dB and the Rician factor is set to 5 dB. The bandwidth is 1 MHz. We also set $\bar{a} = 2.463$, $\bar{b} = 1.635$, and $c=0.826$ for the EH model of all SRs.
\begin{figure}[!t]
\centering
\captionsetup{justification=centering}
\includegraphics[scale = 0.6]{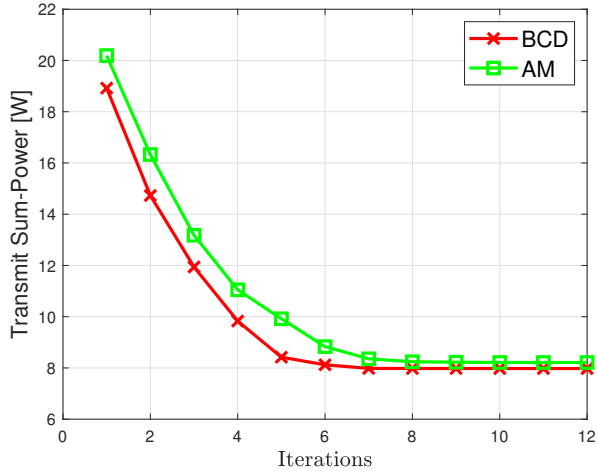}
\caption{Convergence behavior of the AM and BCD algorithms.}
\label{fig:conv}
\end{figure} 

\textit{1) Convergence:} In Fig.~\ref{fig:conv}, we plot the convergence behavior of the AM and BCD algorithms (Alg.~2 and Alg.~3, respectively). We note that they both converge quickly, after less than 10 iterations, as well as that the BCD algorithm converges faster than the AM one.

\begin{figure}[!t]
\centering
\subfloat[]{
	\label{subfig:4a}
	\includegraphics[scale=0.45]{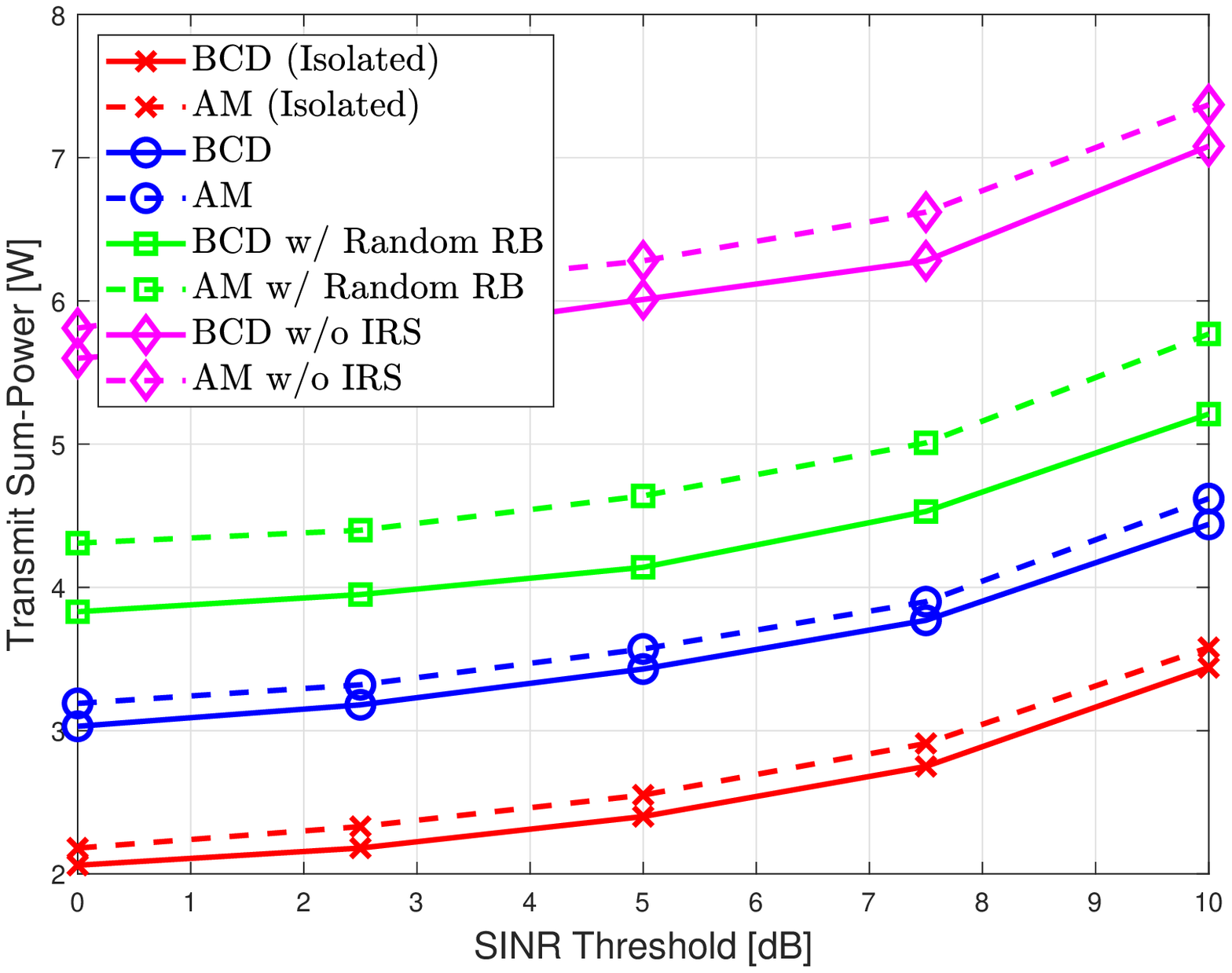} } 
\subfloat[]{
	\label{subfig:4b}
	\includegraphics[scale=0.45]{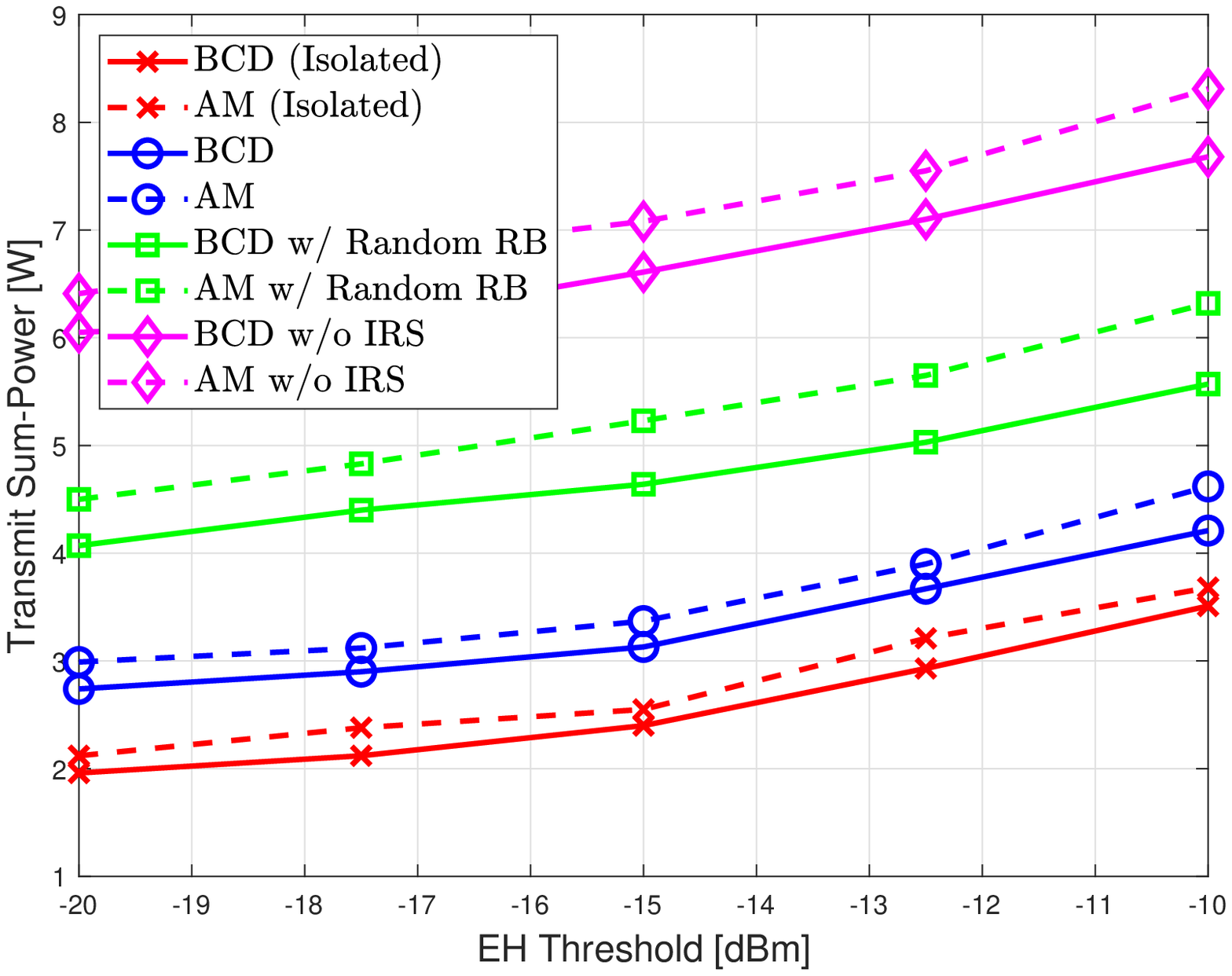} } 
	
\caption{Transmit sum-power for varying: (a) SINR threshold; (b) EH threshold.}
\label{fig:4}
\end{figure}

\textit{2) Impact of the QoS Requirements and the IETs:} We proceed with the test cases that belong to the perfect CSI scenario. First, we plot the transmit sum-power of the ST versus the QoS targets of the SRs. Specifically, in Figs.~\ref{subfig:4a} and~\ref{subfig:4b}, we vary the SINR and EH targets, respectively, while keeping the other parameters fixed. We compare the performance of the proposed AM and BCD algorithms in the considered IRS-aided spectrum underlay scenario with that in the following benchmark scenarios: i) Isolated: There is no primary system, i.e., we considered a stand-alone secondary system. In this case, we solve problem (P1) with the interference power constraints inactive. We also null the RISI terms in the QoS constraints. ii) Random RB: The IRS utilizes random phase shifts. In this case, we solve in each iteration only the problem (P3) with the given phase shifts. iii) No IRS: There is no IRS deployed. In this case, we again solve only problem (P3) in each iteration, but this time we replace the effective transmitter--users channels with the corresponding direct channels.  

We note that the use of the IRS results in significant performance gains, in terms of transmit power minimization. This is true even for the case where the sub-optimal scheme that employs random phase shifts is applied. The performance gap between the IRS-aided system and the equivalent system that does not involve an IRS becomes larger for more demanding QoS requirements, since in this case the impact on the performance of the IRS's ability to enhance the SINR and the received (or, equivalently, the harvested) power at the SRs while mitigating the interference at the PRs is more prominent. We also observe that the transmit power of the stand-alone system is smaller than that of the interference-constrained system. This is because in the latter case, transmit precoding design should comply with the inter-system interference suppression requirements, thus leading to less power-efficient TB directions. Finally, we notice that the BCD algorithm outperforms its AM counterpart. Therefore, in the following tests, we will consider only this algorithm.

\begin{figure}[!t]
\centering
\subfloat[]{
	\label{subfig:5a}
	\includegraphics[scale=0.45]{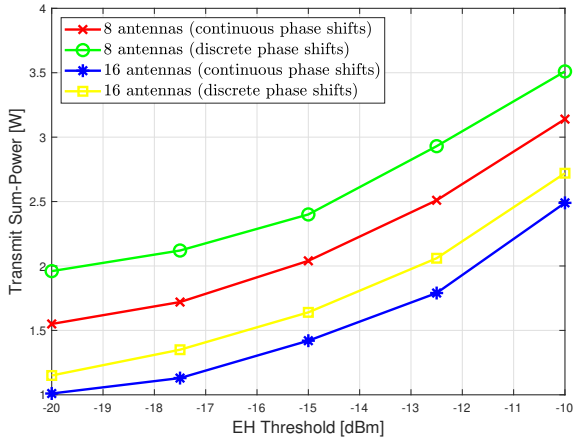} } 
\subfloat[]{
	\label{subfig:5b}
	\includegraphics[scale=0.45]{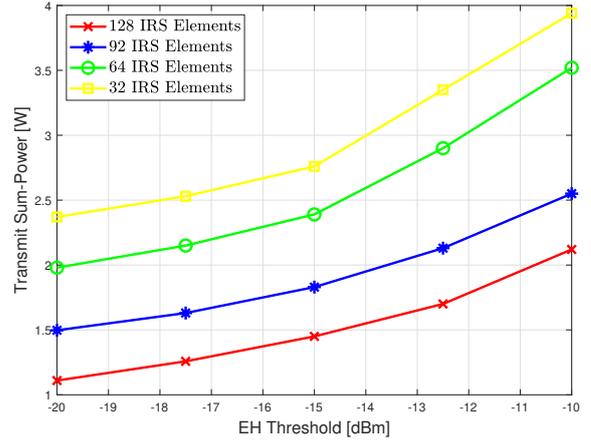} } 
	
\caption{Transmit sum-power for varying number of: (a) ST antennas; (b) IRS elements.}
\label{fig:5}
\end{figure}

\textit{3) Impact of the Number of Transmit Antennas, IRS Elements, Cooperation Overhead, and Discrete IRS Phase Shifts:} In the subsequent test cases, we plot the total transmit power versus the EH threshold for varying number of transmit antennas $M$ or IRS elements $N$. We note in Fig.~\ref{subfig:5a} that the increased spatial degrees-of-freedom offered by a larger number of transmit antennas results in reduced total transmit power. This performance gain is more prominent for more demanding EH targets. Similar conclusions are derived as we are increasing the number of IRS elements, as depicted in Fig.~\ref{subfig:5b}, thanks to the higher corresponding passive beamforming gains. However, the gains are smaller than expected in an ideal case, since as we increase $M$ or $N$, the cooperation overhead and, therefore, the respective performance penalty increases as well. We also notice in Fig.~\ref{subfig:5a} that the mapping of continuous IRS phase shift solutions to the closest discrete values of actual implementations results in a moderate performance loss (which is naturally reduced as the number of discrete levels increases).  

\begin{figure}[!t]
\centering
\subfloat[]{
	\label{subfig:6a}
	\includegraphics[scale=0.45]{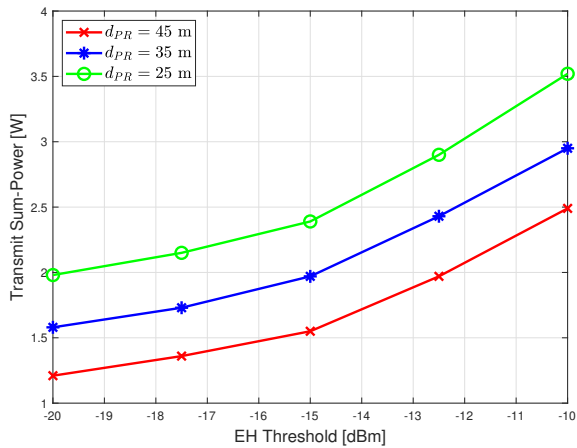} } 
\subfloat[]{
	\label{subfig:6b}
	\includegraphics[scale=0.45]{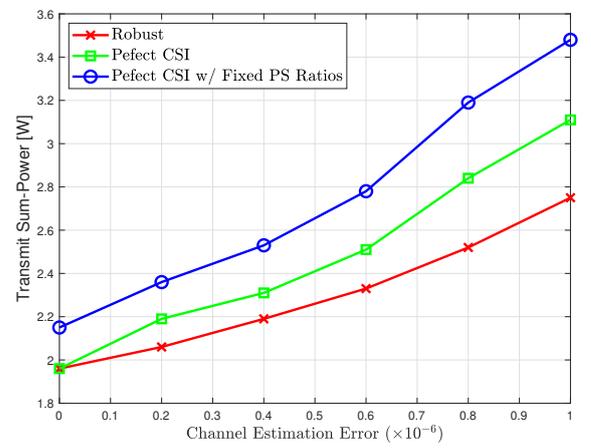} } 
	
\caption{Transmit sum-power for varying: (a) ST--PR distance; (b) channel estimation error.}
\label{fig:6}
\end{figure}

\textit{4) Impact of Node Distances:} In this test case, we plot the total transmit power versus the EH threshold as we vary the distance of the PRs cluster from the ST. We see in Fig.~\ref{subfig:6a} that as we move the PRs closer to the ST, the transmit power required to meet the QoS demands of the SRs and the interference constraints of the PRs increases, due to the increased levels of FISI.

\textit{5) Impact of the Channel Uncertainty and Fixed PS Ratios:} Finally, we plot the total transmit power vs. the channel estimation error for fixed PS ratios, as discussed at the beginning of Sec.~\ref{sec:4}, with $\omega_{R}=0.015$ and $\omega_{E}=1$. We observe in Fig.~\ref{subfig:6b} that the robust scheme presents satisfactory performance, in contrast to the non-robust scheme whose performance is degraded significantly as the channel estimation increases. Finally, with fixed (i.e., non-optimized) PS ratios, we have to waste more transmit power in order to meet the EH requirements of the SRs.

\textit{Summary and design guidelines:} We notice that in order to achieve satisfactory performance, i) we need a large number of IRS elements, and ii) the PRs should not be very close to the ST.

\section{Summary and Conclusions}\label{sec:6}
In this work, we considered an IRS-aided secondary multi-user MISO system for PS-SWIPT that is collocated with a primary multi-user MISO downlink system. We studied the transmit sum-power minimization problem subject to the QoS requirements of the SRs and the IPTs of the PRs. We proposed a side information acquisition protocol and penalty-based algorithms for both the perfect and imperfect CSI case. Numerical simulations highlighted the performance gains of the proposed schemes and provided valuable insights.

\bibliographystyle{unsrt}
\bibliography{refs}
\end{document}